\documentclass[english,aps,twocolumn,prX,showpacs,superscriptaddress]{revtex4-1}
\usepackage[T1]{fontenc}
\usepackage[latin9]{inputenc}
\usepackage{babel}
\usepackage{graphicx}
\usepackage{bm}
\usepackage{array}
\usepackage{amsmath}
\usepackage{amssymb}
\usepackage{dcolumn}
\usepackage{epstopdf}
\usepackage{bbold}

\usepackage{color}

\newcommand{\minline}[2]{$\mbox{\fontsize{#1}{10}\selectfont #2}$}

\begin{document}

\title{\textit{Ab initio} investigation of light-induced relativistic spin-flip effects in  magneto-optics}

\author{Ritwik Mondal}
\email[]{Ritwik.Mondal@physics.uu.se}
\affiliation{Department of Physics and Astronomy, Uppsala University, Box 516, SE-75120 Uppsala, Sweden}
\author{Marco Berritta}
\affiliation{Department of Physics and Astronomy, Uppsala University, Box 516, SE-75120 Uppsala, Sweden}
\author{Karel Carva}
\affiliation{Department of Physics and Astronomy, Uppsala University, Box 516, SE-75120 Uppsala, Sweden}
\affiliation{Charles University, Faculty of Mathematics and Physics, Department
of Condensed Matter Physics, Ke Karlovu 5, CZ-12116 Prague 2, Czech
Republic}
\author{Peter M.\ Oppeneer}
\affiliation{Department of Physics and Astronomy, Uppsala University, Box 516, SE-75120 Uppsala, Sweden}
%\email{peter.oppeneer@physics.uu.se}
\begin{abstract}
Excitation of a metallic ferromagnet such as Ni with an intensive femtosecond laser pulse causes an ultrafast demagnetization within approximately 300 fs. It was proposed that the ultrafast demagnetization measured in femtosecond magneto-optical experiments could be due to relativistic light-induced processes.
% either direct light-induced spin-flip processes or coherent relativistic quantum electrodynamics. 
We perform an \textit{ab initio} investigation of the influence of relativistic effects on the magneto-optical response of Ni. To this end, we  develop, first, a response theory formulation of the additional appearing ultra-relativistic terms in the Foldy-Wouthuysen transformed Dirac Hamiltonian due to the electromagnetic field, and, second, compute the influence of relativistic light-induced spin-flip transitions on the magneto-optics.  Our \textit{ab initio} calculations of relativistic spin-flip optical excitations predict that these can give only a very small contribution  ($\le 0.1$\%) to the laser-induced magnetization change in Ni.
\end{abstract}
\date{\today}

\maketitle
\section{Introduction}

Ultrafast laser-induced demagnetization of metallic ferromagnets was discovered in 1996 by Beaurepaire \textit{et al.} \cite{beaurepaire96}, who observed that a ferromagnetic Ni film could be demagnetized to $\sim$50\% in about 300 femtoseconds after excitation with a short laser pulse.
This surprising discovery was followed by many pump-probe magneto-optical experiments on elemental metallic ferromagnets that confirmed the phenomenon of laser-induced demagnetization (see e.g., Refs.\ 
\cite{hohlfeld97,scholl97,regensburger00,kampfrath02,vankampen05,cheskis05,carley12}).
More recently, ultrafast laser-induced demagnetization has been studied in multi-sublattice or multilayer materials employing element resolved probing techniques in the extreme ultraviolet and soft x-ray regimes \cite{boeglin10,radu11,mathias12,rudolf12,eschenlohr13,bergeard14}.

The discovery of ultrafast laser-induced demagnetization led to an intensive debate on what the underlying microscopic mechanism of the ultrafast dissipation of spin angular momentum could be \cite{bovensiepen09,kirilyuk10,carva11nat}.  Several mechanisms have been proposed to explain the ultrafast demagnetization and these continue to be discussed \cite{zhang00,carpene08,koopmans10,krauss09,battiato10}.  An early microscopic explanation was based on direct transfer of angular momentum from the light involving the spin-orbit interaction \cite{zhang00}.
Another proposed mechanism is spin dissipation through fast Elliott-Yafet electron-phonon spin-flip scatterings \cite{koopmans10}. Other proposals are electron-magnon spin-flip scattering \cite{carpene08} or electron-electron spin-flip scattering \cite{krauss09}. A different scenario is based on the laser-generation of superdiffusive spin currents that transport spin angular momentum out of the  excited ferromagnetic film, thus reducing its net magnetization  \cite{battiato10,battiato12}. Other explanations have focused on the direct action of the laser on the electron's spin, causing either a direct, laser-induced spin-flip \cite{zhang09} or a change of the spin through an ultra-relativistic spin-light interaction \cite{bigot09}.

 Despite the still ongoing debate on the mechanism of ultrafast demagnetization it has been shown that spin dynamics simulations within the Landau-Lifshitz-Gilbert, Landau-Lifshitz-Bloch, or Landau-Lifshitz-Baryakhtar formulations \cite{Atxitia10,Ostler12,Wienholdt13,Baryakhtar13}
can be used to describe ultrafast laser-induced demagnetization in alloys when a sufficiently large and fast dissipation of spin angular momentum is assumed.

To establish accurately how much demagnetization can be caused by one of the aforementioned mechanisms density-functional theory (DFT) based electronic structure calculations are indispensable. 
%Such calculations do however go beyond the state of the art  since, currently, the above mentioned model recur to many phenomenological parameters.  
Recently, DFT-based investigations have been performed for the Elliott-Yafet electron-phonon spin-flip scattering in transition metal ferromagnets \cite{carva11,essert11,carva13,illg13}. The \textit{ab initio} calculations predicted relatively small demagnetization rates;  this gave rise to modified proposals, in which in addition an ultrafast reduction of the exchange splitting needed to be taken into account to explain the observed demagnetization \cite{schellekens13,mueller13}. Another recent computational investigation suggested that a combination of spin-flip electron-phonon and electron-magnon scatterings could explain the measured demagnetizations \cite{haag14}.

A demagnetization scenario involving the direct, relativistic spin-photon interaction was proposed a few years ago \cite{bigot09}. In this proposal ultra-relativistic terms stemming from the Dirac Hamiltonian provide a coupling between the electromagnetic field of the pump laser pulse and the spins of electrons in the material \cite{bigot09,dixit13,hinschberger13}. Model calculations of this mechanism have recently been made for transitions from the $2s$ to $2p$ levels  of a hydrogen atom \cite{vonesch12}. A full \textit{ab initio} investigation of the influence of the relativistic spin-photon interaction on the magnetization and magneto-optical response has not yet been made.

Here, we report an \textit{ab initio} investigation of the influence of the relativistic spin-photon interaction. We present first analytic theory to analyze which terms are the relativistic terms that are involved in the coupling of the spin and photon fields. A notable difference as compared to other recent investigations \cite{bigot09,vonesch12} is the direct consideration of the exchange field in our approach. Also, we investigate the influence of the additional relativistic terms on the response theory equations for the magneto-optical spectrum.  The derived expressions are employed in \textit{ab initio} calculations of the magneto-optical Kerr effect (MOKE) of Ni. Our calculations underline that the influence of relativistic, laser-induced spin-flips is present, but is quite small and can thus not account for the substantial amount demagnetization that is observed in femtosecond pump-probe magneto-optical measurements.

In the following we first provide a derivation of the relativistic spin-photon interaction starting from the Dirac equation (Sec.\ \ref{sec:Dirac} and \ref{sec:FW}). In Sec.\ \ref{sec:implementation}  the corresponding momentum operator is derived and results of \textit{ab initio} calculations for Ni are presented in Sec.\ \ref{sec:results}, and conclusions in Sec.\ \ref{sec:conclusions}.

\section{Theory}

\subsection{The Dirac-Kohn-Sham equation}
\label{sec:Dirac}
To include the relativistic light-spin coupling effects
in the calculation of
the magneto-optical Kerr spectra we consider the 
Dirac-Kohn-Sham (DKS) equation \cite{macdonald79,eschrig99,greiner00}
\begin{equation}
\label{eq:DKS_eq}
H|\psi\rangle=(E-mc^2)\vert\psi\rangle
\end{equation}
with $H$ being the DKS Hamiltonian,
\begin{equation}
H=c\,\bm{\alpha}\cdot\bm{p}+
\left(\bm{\beta}-\mathbb{1}\right)mc^{2}+
V + \mu_B \bm{\beta} \, \bm{ \Sigma}\cdot\bm{B}^{\rm xc}.
\label{eq:DKSH_xc}
\end{equation}
Here $\bm{\Sigma}=\mathbb{1}\otimes\bm{\sigma}$ is the spin operator in the Dirac bi-spinor space, $V$ is the unpolarized Kohn-Sham selfconsistent potential, $\bm{B}^{\rm xc}$ is the spin-polarized part of the exchange-correlation potential in the material, $\bm{p}=-i\hbar\bm{\nabla}$, $\mathbb{1}$ is the $4\times4$ identity matrix, and 
$\mu_B$ is the Bohr magneton, $\mu_B = \frac{e \hbar}{2m}$.
The matrices
\begin{equation}
\bm{\alpha}=\left(\begin{array}{cc}
0 & \bm{\sigma}\\
\bm{\sigma} & 0
\end{array}\right)\quad\mbox{} , \quad{\bm \beta} =\left(\begin{array}{cc}
\bm{1} & 0\\
0 & -\bm{1}
\end{array}\right)\
%quad\mbox{}\quad \bm{\Sigma} =\left(\begin{array}{cc}
%\bm{\sigma} & 0\\
% & \bm{\sigma}
%\end{array}\right)
\nonumber
\end{equation}
are the well known Dirac matrices, with $\bm{\sigma}$ the Pauli spin matrices and $\bm{1}$ the $2\times2$ identity matrix.
The fully relativistic state $| \psi \rangle$  in Eq.\ (\ref{eq:DKS_eq}) 
is the Dirac bi-spinor,
%$\psi$ is the Dirac bi-spinor written as
\[
|\psi\rangle= 
%{ \vert\psi_{+}\rangle \choose \vert\psi_{-}\rangle}
\left(\begin{array}{c}
\vert\psi_{+}\rangle\\
\vert\psi_{-}\rangle
\end{array}\right).
\]\\
%
%By means of the Gordon decomposition
%\cite{greiner00}
%and  neglecting the orbital magnetism \cite{macdonald79,eschrig99}
%Eq.\ (\ref{eq:DKS_H}) can be rewritten as
%\begin{equation}
%H=c\bm{\alpha}\cdot\bm{p}+
%\left(\beta-1\right)mc^{2}+
%V - \mu_B \beta \, \bm{\sigma}\cdot\bm{B}^{\rm xc}
%\label{eq:DKSH_xc}
%\end{equation}
%where $\mu_B$ is the Bohr magneton, $\mu_B = \frac{e \hbar}{2m}$ and $\bm{B}^{\rm xc}$ is the spin-polarized part of the exchange-correlation potential.
An important point to observe is that in the DKS equation the exchange field $\bm{B}^{\rm xc}$ is different from the standard magnetic field, as it obviously acts only on the spin degree of freedom  and does not couple to the orbital 
angular momentum. It is thus not a proper magnetic field and cannot be represented by a vector potential \cite{eschrig99}. 
Therefore it is not included as a vector potential $\bm{A}^{\rm xc}$ in the linear momentum, i.e.\ $\bm{p}-e\bm{A}^{\rm xc}$.
However, to account for an external electromagnetic 
perturbation (for instance being due to a laser source) we introduce the vector potential $\bm{A}(\bm{r},t)$, leading to
\begin{equation}
H=c\bm{\alpha}\cdot\left(\bm{p}-e\bm{A}\right)+\left({\bm \beta}-\mathbb{1}\right)mc^{2}+V +\mu_B {\bm \beta} \, \bm{\Sigma}\cdot \bm{B}^{\rm xc} .
\label{eq:DKSH_xc_pert}
\end{equation}
In Appendix A we show that, in the nonrelativistic limit,  this form of the DKS equation leads to a  Hamiltonian where the external magnetic field 
$\bm{B} (\bm{r}, t) =\bm{\nabla}\times\bm{A} (\bm{r},t) $ couples to the spin $\bm{S}$ $(= \frac{\hbar}{2}\bm{\sigma})$ and orbital angular momentum $\bm{L}$ operators, but the exchange field $\bm{B}^{\rm xc}$
couples \textit{only} to the spin operator.

The aim of this work is to 
investigate the influence of relativistic terms -- that lead to a spin-photon field coupling -- 
on the MOKE spectra.
To elucidate the terms that involve both spin degrees of freedom and the external electromagnetic field we rewrite the Hamiltonian in Eq.\ (\ref{eq:DKSH_xc_pert}), as a semirelativistic expansion in terms of order of \minline{9}{$1/c^2$}. Such rewriting of the DKS equation can be achieved in two ways. 
The small component of the wavefunction $| \psi_{-} \rangle$ can be eliminated exactly, leading to an equation which is fully equivalent to the DKS equation, but for the large component $| \psi_{+} \rangle$ only \cite{kraft95}. This equation can subsequently be expanded in orders of \minline{9}{$1/c^2$}.
Alternatively, one can use the
 Foldy-Wouthuysen (FW) transformation approach \cite{foldy50,greiner00}, however, it needs to be 
extended to the case where an exchange  field $\bm{B}^{\rm xc}$ is present, which was not done before. We note that apart from the exact transformation of Kraft \textit{et al}.\ \cite{kraft95}, a Green function technique was applied by Cr\'epieux and Bruno \cite{crepieux01} to obtain the semirelativistic Hamiltonian. However, they did not start from the DKS Hamiltonian as given in Eq.\ (\ref{eq:DKSH_xc_pert}), but instead added the external magnetic field to the exchange field and did not have a vector potential in the momentum, $\bm{p}-e\bm{A}$ (but only $\bm{p}$). As discussed further below, they obtained several similar terms, yet not all that follow from the FW transformation.

In the following we employ the Foldy-Wouthuysen transformation to derive the Hamiltonian terms that give rise to a spin-photon field coupling.

\subsection{The time-dependent Foldy-Wouthuysen transformation}
\label{sec:FW}
To make a clear distinction between pure non-relativistic Schr\"odinger-like terms and relativistic terms
(up to the order of \minline{9}{$1/c^2$}) we use the 
Foldy-Wouthuysen transformation \cite{greiner00}.
Formally, the time-dependent FW transformation
can be expressed as
\begin{equation}
\label{eq:FW_formal}
H_{{\rm FW}}=e^{iU_{\rm FW}}\left(H-i\hbar\frac{\partial}{\partial t}\right)e^{-iU_{\rm FW}}
\end{equation}
where $\mathrm{e}^{iU_{\rm FW}}$ is a unitary operator
that transforms the DKS Hamiltonian
to a block diagonal form, where each block is $2\times 2$. The four-component Dirac Hamiltonian is diagonalized under the assumption that, at all points in the configuration space, two of the spin components are much smaller than the other two. This assumption is valid if the kinetic and potential energy of the electron is much smaller than the rest mass energy of the electron.
Since we are interested only in the ``positive energy'' solutions we will retain only the upper $2\times 2$ component of the Hamiltonian (the large component of the Dirac bi-spinor).
To find the transformed Hamiltonian in a \minline{9}{$1/c^2$} expansion
we write the DKS Hamiltonian as
\begin{equation}
\label{eq:odd_even_separation}
H=\left({\bm \beta}-\mathbb{1}\right)mc^{2}+\mathcal{O}+\mathcal{E}
\end{equation}
with $\mathcal{O}=c\, \bm{\alpha}\cdot\left(\bm{p}-e\bm{A}\right)$ as an odd operator (i.e., off-diagonal in the particle-antiparticle Hilbert space) and $\mathcal{E}=V  + \mu_B {\bm \beta} \, \bm {\Sigma}\cdot\bm{B}^{\rm xc}$ as an even operator (diagonal in the same space).
Next, we consider the FW operator,
\begin{equation}
\label{eq:FW_operator}
U_{\rm FW}=-\frac{i}{2mc^{2}}{\bm \beta} \mathcal{O}.
\end{equation}
To obtain an expansion in orders of \minline{9}{$1/c^2$}
we use a Taylor expansion
of the operator 
%$\mathrm
${e}^{\pm iU_{\rm FW}}\simeq 1\pm iU_{\rm FW}+{O}(\minline{9}{$1/c^2$})$.
This leads to a transformed Hamiltonian
\begin{equation}
\label{eq:hprime}
H^{\prime}=\left({\bm \beta}-\mathbb{1}\right)mc^{2}+\mathcal{O}^{\prime}+\mathcal{E}^{\prime} ,
\end{equation}
where $\mathcal{O}^{\prime}$ is the transformed odd part which is of the order of \minline{9}{$1/c^2$} and $\mathcal{E}^{\prime}$ is the transformed even part.
Repeating  this procedure two times
with the operators $U_{\rm FW}^{\prime}=-\frac{i}{2mc^{2}}{\bm \beta} \mathcal{O}^{\prime}$ to get a further transformed $H^{\prime\prime}$,
$\mathcal{O}^{\prime\prime}$, and $\mathcal{E}^{\prime\prime}$
and then working with the operator
$U_{\rm FW}^{\prime\prime}=-\frac{i}
{2mc^{2}}{\bm \beta}\mathcal{O}^{\prime\prime}$ on the transformed Hamiltonian $H^{\prime\prime}$
we get rid of the odd terms up to the order of \minline{9}{$1/c^6$}.
After these transformations we obtain a
transformed Hamiltonian written in terms of the original odd and even parts
%\begin{widetext}
\begin{eqnarray}
\label{eq:FW_ham_for}
H^{\prime\prime\prime}_{\rm FW} &=& \left({\bm \beta}-\mathbb{1}\right)mc^{2}+{\bm \beta}\left(\frac{\mathcal{O}
^{2}}{2mc^{2}}-\frac{\mathcal{O}^{4}}{8m^{3}c^{6}}\right) \nonumber \\
&+& \mathcal{E}-\frac{1}{8m^{2}c^{4}}[\mathcal{O},[\mathcal{O},\mathcal{E}]+i\dot{\mathcal{O}}]\label{fw} .
\end{eqnarray}
%\end{widetext}
Substituting the explicit form of the operators $\mathcal{O}$ and $\mathcal{E}$, retaining only the terms up to the order \minline{9}{$1/{c^2}$} and keeping 
in mind that, for the vector potential of the external electromagnetic field, $\bm{B}=\bm{\nabla}\times\bm{A}$ and 
$\bm{E}=-\frac{\partial \bm{A}}{\partial 
t}$, we arrive at the Hamiltonian restricted to the large component of the Dirac bi-spinor,
\begin{widetext}
\begin{eqnarray}
\label{eq:FW_hamiltonian}
\!\! H_{\rm FW}&=&\frac{\left(\bm{p}-e\bm{A}\right)^{2}}{2m}+V - \mu_B \,\bm{\sigma}\cdot \bm{B}^{\rm xc}-  \mu_B \,\bm{\sigma}\cdot \bm{B} -\frac{\left(\bm{p}-e\bm{A}\right)^{4}}{8m^{3}c^{2}}-\frac{1}{8m^{2}c^{2}}\left(p^{2}V\right)-
\frac{e\hbar^{2}}{8m^{2}c^{2}}\bm{\nabla}\cdot\bm{E}\nonumber\\
&& +\frac{i}{4m^{2}c^{2}}\bm{\sigma}\cdot\left(\bm{p}V\right)\times\left(\bm{p}-e\bm{A}\right)-\frac{e\hbar}{8m^{2}c^{2}}\bm{\sigma}\cdot\left\{ \bm{E}\times\left(\bm{p}-e\bm{A}\right)-\left(\bm{p}-e\bm{A}\right)\times\bm{E}\right\}\nonumber\\
&&   +\frac{\mu_B}{8m^{2}c^{2}}\Big\{\left[p^{2}(\bm{\sigma}\cdot\bm{B}^{\rm xc})\right]+2\bm{\sigma}\cdot(\bm{p}\bm{B}^{\rm xc})\cdot\left(\bm{p}-e\bm{A}\right)+2(\bm{p}\cdot\bm{B}^{\rm xc})\, \bm{\sigma}\cdot\left(\bm{p}-e\bm{A}\right)+4[ \bm{B}^{\rm xc}\cdot\left(\bm{p}-e\bm{A}\right)]\, \bm{\sigma}\cdot\left(\bm{p}-e\bm{A}\right)\Big\} \nonumber\\
&& +\frac{i \mu_B}{4 m^{2}c^{2}}[(\bm{p}\times\bm{B}^{\rm xc})\cdot\left(\bm{p}-e\bm{A}\right)] .
\end{eqnarray}
\end{widetext}
Note that, when the momentum operator and a (vector) function are enclosed in round brackets the momentum operator acts only on this function. This Hamiltonian is an extension to the conventional Pauli Hamiltonian (see Appendix \ref{sec:pauli}) yet, including all the \minline{9}{$1/c^2$} terms and all the terms involving $\bm{B}^{\rm xc}$ to the same order. We note in addition that the Hamiltonian (\ref{eq:FW_hamiltonian}) is quite different 
 from the Hamiltonian given by Bigot \textit{et al.} \cite{bigot09}, as  they did not consider the magnetic exchange interaction, which however is the strongest magnetic interaction in a ferromagnetic material as Fe, Co, or Ni. In a further work Vonesch and Bigot \cite{vonesch12} considered a static homogeneous applied magnetic field, expressed by a vector potential, as well as a time-varying vector potential to describe the electromagnetic field. Such static homogeneous magnetic field is nonetheless different from the exchange field $\bm{B}^{\rm xc}$, as the latter, as mentioned in the previous section, cannot be included by means of a vector potential \cite{macdonald79,eschrig99}. Specifically, compared to the extended Pauli Hamiltonian (\ref{eq:FW_hamiltonian}), Vonesch and Bigot obtain the first, second, fourth, sixth, and eighth terms, as well as a term similar (but not identical) to the ninth term.  As they didn't include an exchange field, they did not obtain any  of the terms containing $\bm{B}^{\rm xc}$, but had a  
 contribution $\bm{\sigma}\cdot \bm{B}_{\rm ext}$ due to the static applied homogeneous field. In addition they found an additional term, $\bm{A}_L \cdot \bm{A}_{\rm ext}$, the product of the two vector potentials of the electromagnetic radiation and the constant external magnetic field. This term, which stems from writing out $\left(\bm{p}-e[\bm{A}_L + \bm{A}_{\rm ext} ] \right)^{2}$, does not appear in our formulation where the exchange field is not represented by a vector potential.
 
 Our extended Pauli Hamiltonian (\ref{eq:FW_hamiltonian}) can further be compared with the Hamiltonian obtained by Cr\'epieux and Bruno \cite{crepieux01} using a Green's function-based method. 
%In ref.\cite{crepieux01} it had been already addressed the problem of finding, with a method involving Green's function, the $1/c^2$ limit of the Driac equation. 
%Here the authors 
They considered two different Hamiltonians, one with an effective (including exchange) vector potential $\bm{A}_{\rm eff}$, and the other one with an effective magnetic field, $\bm{B}_{\rm eff}$. In the first case they obtained the same terms as we do from the introduction of the external vector potential, however, the feasibility of the DFT-based formulation of such Hamiltonian appears to be an open question. In the second case, they apply their method to an effective Hamiltonian where the orbital magnetic effects are neglected, and make the critical assumption $\bm{B}_{\rm eff}=\bm{\nabla}\times\bm{A}_{\rm eff}$; however, on account of its nature, the exchange part of their $\bm{B}_{\rm eff}$ does not satisfy the Maxwell equations as a normal magnetic field  does (see e.g.\ Ref.\ \cite{eschrig99}). In our case we used a vector potential to account for an external field without any need to use the mentioned critical assumption, %above-cited for our additional $B_{xc}$, this reflecting the appearance, i
thus giving in our Hamiltonian the proper $\bm{E}$-dependent terms and terms linear in both, the external vector potential and $\bm{B}^{\rm xc}$ fields, which are missing in the approach of Ref.\ \cite{crepieux01}.
%Comparing our Hamiltonian in Eq. (\ref{eq:DKSH_xc_pert}) with both of the effective Hamiltonian, they considered $\bm{A}_{\rm eff}$ as the same spirit of $\bm{A}(\bm{r},t)$ and $\bm{B}_{\rm eff}$ as the same spirit of $\bm{B}^{\rm xc}$ in our case. Thus we get explicit dependence of terms on exchange field and external electromagnetic field. Moreover, we get exactly the same $1/c^2$ terms involving $\bm{B}^{\rm xc}$ as it involves $\bm{B}_{\rm eff}$ in their work.-M

The Hamiltonian (\ref{eq:FW_hamiltonian}) looks cumbersome at first sight
but its physical content is readily explained.
\begin{itemize}
\item The first and second terms comprise the 
usual Schr\"odinger Hamiltonian
for a particle in an external field $V$ (which in this case is a selfconsistent potential) and 
where the minimal coupling with an external vector potential $\bm{A}(\bm{r},t)$ is present (this may represent, in general, any kind of external electromagnetic field).
\item The third term is a Zeeman-like term due to the presence of the magnetic exchange field.
\item The fourth term is the standard Zeeman term with the external magnetic field.
\item The fifth term is the relativistic mass correction.
\item The sixth and the seventh terms are respectively the Darwin terms related with the selfconsistent potential $V$ and the standard Darwin term arising from the external perturbation.
\item The eighth and the ninth terms are those which in a central potential $V$ give rise to the spin-orbit coupling. 
\item All the remaining terms, except the last, can be seen as corrections to the spin-orbit coupling due to the spin-polarized exchange field.
This is more apparent using the identity 
\begin{equation}
r \bm{p}=\bm{r}\hat{p}_{r}-
\frac{\bm{r}}{r}\times\bm{L} ,
\end{equation}
where 
%$\bm{L}$ is the angular momentum operator and 
$\hat{p}_r=-i\hbar\frac{\partial}{\partial r}$ is the spatial part of the momentum operator.
\item The last term depends on the $\bm{B}^{\rm xc}$ field
but is independent of the spin.
%\item \textcolor{red}{We still have to understand the last term}.
\end{itemize}
The terms which involve a direct coupling of the spin to the external electromagnetic field are the fourth, eighth, ninth, and tenth ones. 
Via these terms it would in principle be possible to control the spin of electrons in a magnetic material by applying an external laser field.\\

\subsection{Strategy for the numerical implementation}
\label{sec:implementation}
Our aim is to implement the above-derived relativistic terms in a suitable formalism for \textit{ab initio} calculations. An adequate way to achieve this is to consider the change of the Hamiltonian due to the applied electromagnetic field, which then is treated as a perturbation within Kubo linear-response theory to obtain the corresponding optical conductivity tensor (see, e.g.\ \cite{oppeneer01}).

The standard strategy for the derivation of the conductivity tensor consists in gathering the external magnetic vector potential related linear terms in an interaction Hamiltonian \cite{landau81},
\begin{equation}
\delta\langle H_I\rangle=-\int \! d\bm{r} \,\bm{j}\cdot\delta\bm{A},
\label{eq:H_int}
\end{equation}
and then to
rewrite the current density operator $\bm{j}$ in terms of the momentum operator.
This procedure is straightforward in the fully relativistic approach [see Eq.\ (\ref{eq:DKSH_xc_pert})]
because the momentum operator, which is the conjugate of the position operator, is given as
\begin{equation}
\label{eq:conj_DKS}
\bm{\Pi}_{\rm D}=-\frac{im}{\hbar}[\bm{r},H]=mc\,\bm{\alpha} ,
\end{equation}
and the variation of the DKS Hamiltonian [Eq.\ (\ref{eq:DKSH_xc_pert})] results easily
\begin{equation}
\label{eq:var_DKS}
\bm{j}=-\frac{\delta H_I}{\delta \bm{A}}=ec\, \bm{\alpha} = \frac{e}{m}\bm{\Pi}_{\rm D}.
\end{equation}
However, in the semirelativistic limit this equivalence breaks down.
The expression of the conjugate momentum operator, obtained 
using the position-momentum conjugation relation starting from the extended Pauli Hamiltonian
[Eq.\ (\ref{eq:FW_hamiltonian})], is:
\begin{widetext}
\begin{equation}
\label{eq:rel_mom}
\bm{\Pi}_{\rm P}=
%\frac{im}{\hbar}[H,\bm{r}]=
\bm{p}+\frac{1}{4mc^{2}}\left[\frac{2p^{2}}{m}\bm{p}+i\bm{\sigma}\times(\bm{p}V)+ \mu_B \Big\{ \bm{\sigma}\cdot(\bm{p}\bm{B}^{\rm xc})+(\bm{p}\cdot\bm{B}^{\rm xc})\bm{\sigma}+2\bm{B}^{\rm xc}(\bm{\sigma}\cdot\bm{p})+2\bm{\sigma}(\bm{B}^{\rm xc}\cdot\bm{p})+i(\bm{p}\times\bm{B}^{\rm xc}) \Big\} \right]
\end{equation}
\end{widetext}
where all the terms in the square brackets are due to relativistic corrections.  The first and the second terms in the square brackets are related  to the relativistic mass correction and the spin-orbit coupling, respectively, and can be obtained by means of the standard FW transformation in the absence of the exchange field; all the remaining terms are new and stem from relativistic corrections and the exchange field in the DKS equation. To reformulate this momentum operator, we use that 
it has been seen in Ref.\ \cite{nowakowski99} that the current density operator in the semirelativistic limit can be written in the form
\begin{equation}
\bm{j}(\bm{r})=\bm{j}_{\pi}+\bm{j}_{sp}= \frac{e}{m}\bm{\Pi}_{\textrm P} -\frac{ie}{2m}(\overleftarrow{\bm{p}}\times 
\bm{\sigma}+ \bm{\sigma}\times{\bm{p}})
\end{equation}
where $\overleftarrow{\bm p}$ means that the operator $\bm{p}$ operates to the left. Here the second and third terms, representing $\bm{j}_{sp}$, the spin-polarized current density,
%after the last equality of this expression 
underline a crucial difference between the momentum operator and the current density operator.
These terms can be obtained in several ways.
They can be derived from a variation of the Hamiltonian with respect to $\bm{A}$ as already done in 
\cite{landau81}, or they can be derived by applying 
the Foldy-Wouthuysen transformation to the fully relativistic Dirac current $\bm{j}=ec\,\bm{\alpha}$ \cite{huszar67}. Here, we adopted a different approach to derive the spin-polarized current, see Appendix \ref{zeeman_current}.
For our purpose, however, it is important to note that the additional $\bm{j}_{sp}$ term
does not give any contribution. In fact, its matrix elements which are required for an \textit{ab initio} calculation, are
\begin{eqnarray}
\label{eq:spin_pol_curr}
\!\!\!\! \!\!
\langle\Psi_n\vert\bm{j}_{sp}\vert\Psi_m\rangle&=&  \frac{e \hbar}{2m}  \int_{V}
d\bm{r} \,
\bm{\nabla}\times\left[\Psi_n^{*}(\bm{r})
\bm{\sigma}\Psi_m(\bm{r})\right]
\nonumber\\
&=&   \frac{e \hbar}{2m} \int_{S}
d\Omega \,
\bm{n}\times\left[\Psi_n^{*}(\bm{r})
\bm{\sigma}\Psi_m(\bm{r})\right] ,
\end{eqnarray}
 with $\bm{n}$ the unit vector normal to the surface.
This integral vanishes when we consider the  infinite surface of  integration that encloses our sample.
Consequently, we have proved that, for our purpose, we can work with the current operator given by the momentum operator $\bm{\Pi}_{\rm P}$.
%we can substitute in the linear response in eq.\ (\ref{eq:conductivity}) 

\subsection{Optical conductivity tensor and MOKE}
\label{sec:MOKE}

Once the required current density operator, and thus the interaction Hamiltonian Eq.\ (\ref{eq:H_int}),  is known, linear-response theory can be applied to obtain the conductivity tensor  $\sigma_{\alpha \beta}(\omega)$.
The most accurate way to evaluate the conductivity tensor would be to invoke the linear-response time-dependent DFT which includes the contribution of the exchange kernel to account for non-equilibrium electron interaction processes in the excited state  \cite{ullrich11,marques12}.  However, such calculations on the here-desired relativistic level have not yet been performed. Also, it is our aim   to interpret MOKE experiments where the employed laser intensity is less then $10^{11}$ W/m$^2$. In this regime the number of electrons removed from below the Fermi energy in Ni is about 0.01 electron \cite{koopmans00} and therefore we do not expect that such rearrangement of electrons will give rise to a significant contribution from excited-state electron-electron interaction. Furthermore,  it was shown previously that the bare Kohn-Sham linear-response theory described very well the measured magneto-optical spectra of metals \cite{oppeneer01}.
%the bare Kohn-Sham linear response theory proved to describe very well the linear response regime \cite{oppeneer01} and the non-equilibrium dynamics plays a marginal role in the prediction of the MOKE \cite{oppeneer04}} 

As it is shown in Appendix \ref{sec:lin_resp} the conductivity tensor within the Kohn-Sham linear-response theory can be expressed by an identical equation valid for the fully relativistic, semirelativistic and nonrelativistic  ($\bm{j} = \frac{e}{m} \bm{p}$) cases. In
terms of the Kohn-Sham single-particle energy dispersion relations $\epsilon_{n}(\bm{k})$ and matrix elements of the current operator $j_{nm}^{\alpha}(\bm{k})$, which can be easily obtained in the framework of  band structure calculations, it reads
\begin{eqnarray}
\label{eq:conductivity}
\sigma_{\alpha\beta}(\omega)\approx-\frac{i}{\hbar V}& {\displaystyle \sum \limits_{nn^{\prime}\bm{k}}}&\Big[ \frac{f(\epsilon_n(\bm{k}))-f(\epsilon_{n'}(\bm{k}))}{\omega_{nn'}(\bm{k})}\times\nonumber\\
&&\frac{j^{\alpha}_{n'n}(\bm{k})j^{\beta}_{nn'}(\bm{k})}{\omega-\omega_{nn^{\prime}}(\bm{k})+{i}/{\tau}}\Big] ,
\end{eqnarray}
where $\hbar \omega_{nn^{\prime}}(\bm{k})=\epsilon_{n}(\bm{k})-\epsilon_{n^{\prime}}(\bm{k})$.
Here the parameter $\tau$ accounts for the lifetime broadening. 
%Notably, this expression is valid for the fully relativistic, quasirelativistic and nonrelativistic  ($\bm{j} = -\frac{e}{m} \bm{p}$) cases.

It has already been proven that magneto-optical Kerr spectra are well described in a DFT  band-structure
framework using this single-particle formulation of the linear-response theory \cite{oppeneer1992ab,kraft95}.
 The polar MOKE spectra are related to the conductivity tensor elements
by the equation
\begin{eqnarray}
\label{eq:cond_kerr}
\Phi_{\rm K}(\omega)&=&\theta_{\rm K}(\omega)+i {\varepsilon}_{\rm K}(\omega)\nonumber\\&\simeq&
-\frac{\sigma_{xy}(\omega)}{\sigma_{xx}(\omega)\sqrt{1+\frac{4\pi i}{\omega}\sigma_{xx}(\omega)}}.
\end{eqnarray}
Here the exchange field is chosen along the local $z$ axis and $\Phi_{\rm{K}}$ is the complex polar Kerr angle that can be divided in the real Kerr rotation $\theta_{\rm{K}}$ and imaginary Kerr ellipticity $\varepsilon_{\rm{K}}$. Expression (\ref{eq:cond_kerr}) is not exact, but in our actual calculations below we used the longer exact expression \cite{oppeneer01}.
 Note that in pump-probe magneto-optical experiments the transient MOKE signal is measured, which is often taken to be a direct measure of the changing atomic magnetization \cite{beaurepaire96,zhang09}.

In the following we compute the influence of the relativistic spin-photon couplings terms on the magneto-optical spectra of Ni. 
To evaluate their influence, all the calculations described in the next section are performed with
switching on and off the terms in square brackets of Eq.\ (\ref{eq:rel_mom}). To be precise our \textit{ab initio} implementation
of the relativistic momentum operator matrix elements use the equation (10) in Ref.\ \cite{kraft95} which is
exact to all order in \minline{9}{$1/c^2$}.

\section{Magneto-optical Kerr effect calculations for nickel}
\label{sec:results}

To elucidate the influence of the relativistic spin-photon interaction on the magneto-optical response,
we preformed  \textit{ab initio} calculations for nickel.
In particular we compare computed Kerr effect spectra
 obtained using either the nonrelativistic expression of the momentum operator 
$\bm{p}=-i\hbar\bm{\nabla}$ in the evaluation of the conductivity tensor Eq.\ (\ref{eq:conductivity}), which determines the 
polar Kerr effect, or the semirelativistic expression
for the momentum operator in 
Eq.\ (\ref{eq:rel_mom}).

 It is already well known that the magneto-optical Kerr effect is by itself
a relativistic effect, since it is directly related to spin-orbit coupling \cite{Argyres55,oppeneer1992abso} present in the \textit{ab initio} calculated electronic structure (particularly, in the wavefunctions). The latter we compute with a relativistic (four-component) extension of the augmented-spherical wave (ASW) code
\cite{williams79}, adopting the local spin density approximation (LSDA) to the DFT. 
As has been shown previously, using only the nonrelativistic momentum operator in the conductivity tensor calculations (in conjunction with fully relativistic electronic structure calculations) provides a good description of the MOKE of metallic ferromagnets \cite{oppeneer01}, including Ni. 

%In Ref.\ \cite{oppeneer1992abso} {\em ab-initio} calculations have been performed varying the strength of the spin orbit coupling multiplying it by a factor $\alpha$ ($H_{S-O}=\xi\mathbf{L}\cdot\mathbf{S}\rightarrow\alpha\xi\mathbf{L}\cdot\mathbf{S}$) in the {\em ab-initio} calculations of the band structures and of the eigenstates to be used in the conductivity calculations. The authors found that, for $\alpha$ varying between $0.5-2.0$, the MOKE is linear in $\alpha$. However in the above-cited work it has been used only the non-relativistic momentum operator in the conductivity tensor calculations.\\
%MOKE spectra calculations including all the relativistic terms also in the conductivity tensor calculations, together with the results in Ref.\cite{zhang09} where the controversy about the reliability of the time resolved MOKE (TRMOKE) spectroscpopy was solved (at least for laser pulses duration $\ge70fs$), would give a full reliability to the TRMOKE experiments and their theoretical interpretation.
%For this reasons in this work, -M} 

\begin{figure}[tb]
\centering
\includegraphics[width=0.45\textwidth]{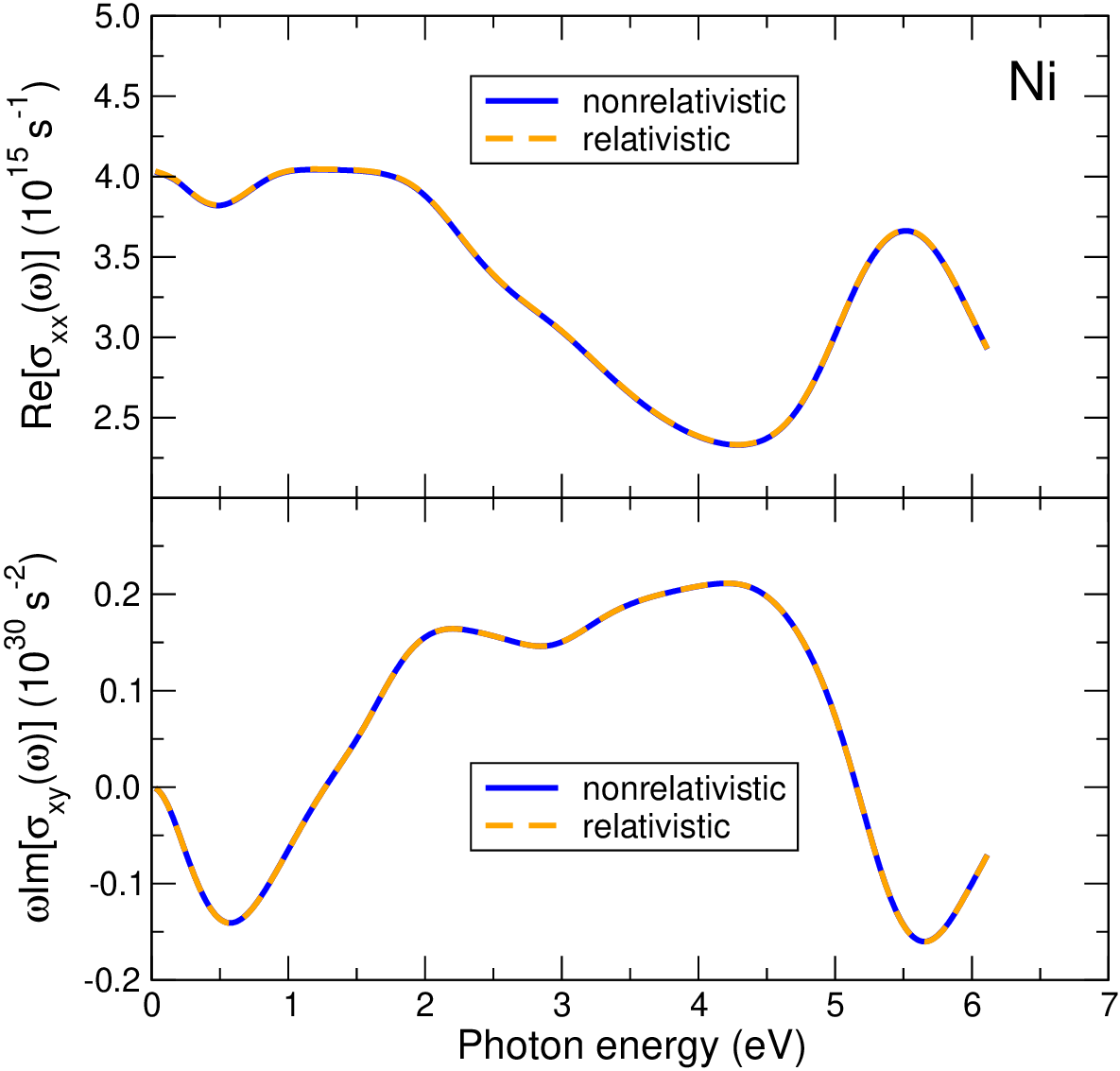}
\caption{
\label{fig:sigxx}
(Color online) Calculated optical conductivity spectra ${\rm Re}[\sigma_{xx}(\omega)]$ (top panel) and $\omega {\rm Im}[\sigma_{xy}(\omega)]$ (bottom panel) of fcc Ni, using the relativistic and nonrelativistic current density formulations.
The blue lines are calculated considering the momentum operator in nonrelativistic approximation ($\bm{p}=-i\hbar\bm{\nabla}$).
The orange dotted-dashed lines are the calculations performed using Eq.\ (\ref{eq:rel_mom}) as momentum operator.
}
\end{figure}
\begin{figure}[tb]
\includegraphics[width=0.41\textwidth]{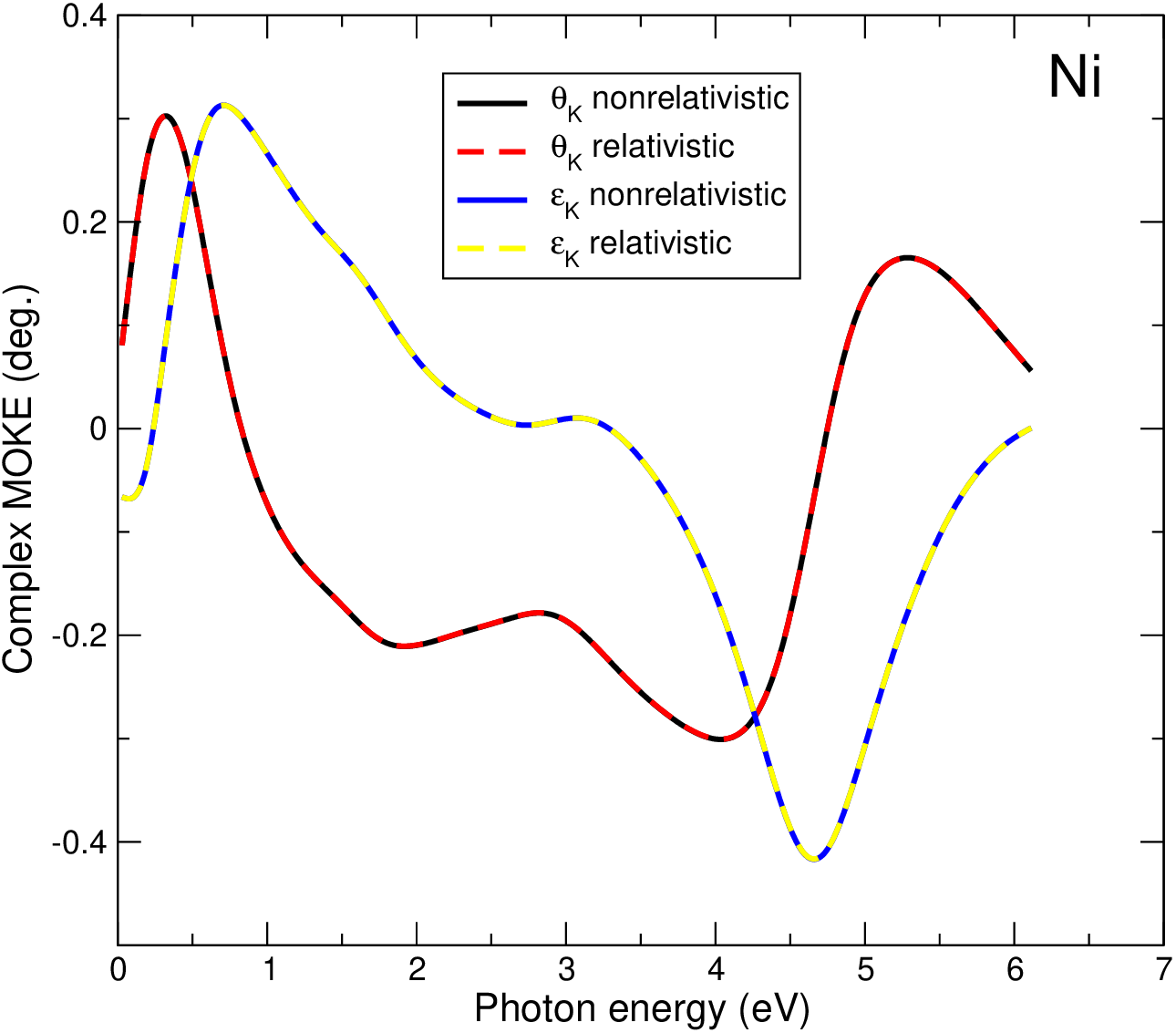}
\caption{
\label{fig:rel_non-rel_comp}
(Color online) Calculated magneto-optical Kerr rotation and Kerr ellipticity of Ni.
The black and blue lines are, respectively, the Kerr rotation $\theta_{\rm K}$ and ellipticity $\varepsilon_{\rm K}$ calculated without taking into account the additional relativistic terms of the momentum operator. The red and yellow dashed lines are  the Kerr rotation and ellipticity, respectively,
calculated retaining the relativistic terms of the momentum operator.
\\}
\end{figure}

In Fig.\ \ref{fig:sigxx} we show 
the comparison between the interband-only optical conductivity elements,
${\rm Re}[\sigma_{xx}(\omega)]$ and 
$\omega{\rm Im}[\sigma_{xy}(\omega)]$, computed with the 
nonrelativistic momentum operator as well as with
including the relativistic corrections to the 
momentum operator. The calculations are
performed with a broadening $\hbar/\tau =0.03$\,Ry. 
As it is apparent from the plot the 
contribution due to the 
relativistic terms does not lead to an
appreciable change in the conductivity spectra.
The influence of the additional light-spin interaction terms on the MOKE spectra is shown in Fig.\ \ref{fig:rel_non-rel_comp}.
As expected from the results shown in 
Fig.\ \ref{fig:sigxx}, the comparison in
Fig.\ \ref{fig:rel_non-rel_comp} 
shows that, also for 
the Kerr spectra,
the contribution from the relativistic correction terms is very small.

\begin{figure}[h!]
\includegraphics[width=0.45\textwidth]{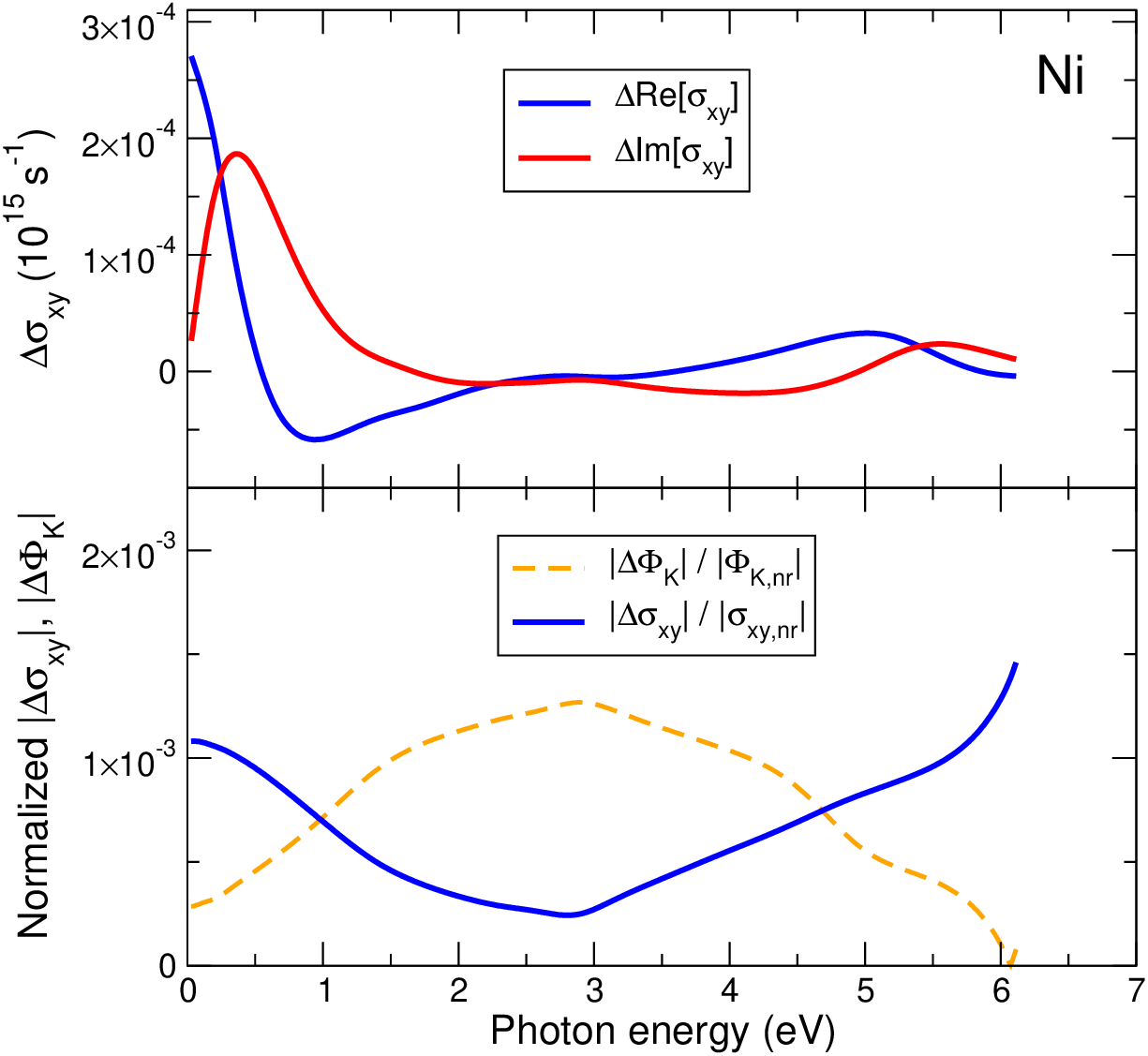}
\caption{
\label{fig:diff} (Color online) Top panel:  differences between the off-diagonal conductivity $\sigma_{xy} (\omega )$ components computed with the relativistic  and the nonrelativistic current operators.
Bottom panel: calculated  absolute value of the difference in $\sigma_{xy}$ ($\Phi_{\rm K}$) normalized to the quantities  $|\sigma_{xy,  {\rm nr}}|$ ($|\Phi_{\rm K,nr}| $) computed with the nonrelativistic momentum operator.
}
\end{figure}

 To quantify the influence of the relativistic terms on the conductivity tensor we show in Fig.\ \ref{fig:diff} (top panel) the difference between the off-diagonal components of the conductivity tensor calculated with and without the relativistic terms in the momentum operator. 
This difference is of the order of $10^{-4}$ ($\times 10^{15}$ s$^{-1}$) for both the real and imaginary part. Note that the difference curves in Fig.\ \ref{fig:diff} are very smooth and do not show any jitter. The reason for its absence is the high numerical accuracy in the calculation of $\sigma_{xy} (\omega )$ which is still appreciably higher (better than $10^{-7}$) (for details of the numerical implementation, see \cite{oppeneer1992ab}). Hence, there is no doubt that we can adequately capture the influence of the relativistic corrections terms. 
In Fig.\ \ref{fig:diff} (bottom panel) we plot the absolute values of the differences in $\sigma_{xy}$  normalized to the nonrelativistic off-diagonal conductivity, i.e.,  $|\Delta\sigma_{xy}|/|\sigma_{xy, \rm{nr}}|$, and the same for the complex Kerr angle, i.e., $|\Delta \Phi_{\rm K}| / | \Phi_{\rm K, nr}|=$ $(\Delta\theta_{\rm K}^2+\Delta\varepsilon_{\rm K}^2)^{1/2}/ (\theta_{\rm K,nr}^2+\varepsilon_{\rm K,nr}^2)^{1/2}$. For both quantities the normalized differences are of the order of 0.1\%. Thus, we can conclude that the relativistic spin-photon terms do contribute to the magneto-optical signal of Ni, but that this contribution is rather small.
 The spin-photon induced change in the MOKE signal would consequently be present during the pump pulse, where it would be interpreted as as magnetization change of the order of 0.1\%.

%\begin{figure}[h!]
%\includegraphics[width=0.45\textwidth]{rel_err.eps}
%\caption{\label{fig:rel_err} Here the expressions $\sqrt{\Delta\theta_k^2+\Delta\varepsilon_k^2}/\sqrt{\theta_{nr_k}^2+\varepsilon_{nr_k}^2}$ (red curve) and $|\Delta\sigma_{xy}|/|\sigma_{xy}|$ (blue curve) are plotted.}
%\end{figure}

%
%Despite this results may sound in contrast
%with the statement that Kerr
%effect has relativistic origin \cite{argyres55,kraft95} it is worth stressing the point that the relativistic effect are fully included in our
%band structure calculations.
%Moreover it has been showed in Ref.\ \cite{oppeneer1992abso} that, modification of the strength of the spin-orbit coupling,
%reflects in proportional changes in the magneto-optical Kerr spectra. {\blue the above part can be said first, perhaps with some further technical details. The computed Kerr effect at 1.55 eV is close to the exp. values. Discuss also Zhang et al NatPhys, say that we have the relevant terms.}
%{\green An other long standing controversy about the MOKE is whether it can be used in time resolved spectroscopy (TRMOKE) Zhang et al \cite{zhang09} performed {\em ab-initio} calculations combined with Liouville equation integration in order to remove the controversy about the real information provided by the time resolved MOKE spectroscopy (TRMOKE). They proved that such shift is due to the temporal profile of the laser pulse that induce dephasing of the signal. They also proved that TRMOKE remain a good spectroscopic strategy unless a laser pulse shorter of $\simeq70fs$ is used. -M}
\section{Discussion and Conclusions}
\label{sec:conclusions}

 Previous estimations of the influence of the relativistic spin-photon interaction have been made by Vonesch and Bigot \cite{vonesch12}, who considered optical transitions on a hydrogen atom within the framework of an extended Pauli Hamiltonian. Calculating the matrix elements of the spin-photon terms in their Hamiltonian, they found that these were of the order of $1 \times 10^{-6}$ (whereas the nonrelativistic term $\bm{p} \cdot \bm{A}$ was of order 1). The largest contribution in their treatment originated from the cross term $\bm{A}_L \cdot \bm{A}_{\rm ext}$ of the vector potential of the laser radiation and an external magnetic field, which, as mentioned before, does not arise in our treatment. Vonesch and Bigot estimated a change in the normalized Kerr rotation of $0.5 \times 10^{-3}$. Thus, even with the additional term this estimated change in the Kerr rotation is in overall accord with ours for metallic Ni. 

 Nonetheless, in spite of the nonzero influence, our \textit{ab initio} calculations do not evidence that relativistic light-induced spin-flip transitions could provide a notable demagnetization channel.  They would appear as a small demagnetization effect during the pump pulse which is in experiments typically about 70\,fs wide. However, during and immediately after the pump pulse there will also be the influence of ``bleaching'', that is, the reduction in the  optical excitation channels caused by the presence  of pump-laser excited electrons \cite{koopmans00,guidoni02}. The influence of such nonequilibrium electron populations on the MOKE spectra of Ni have been evaluated previously \cite{oppeneer04}, yet \textit{without} the here-investigated relativistic spin-photon effects, and were found to be significant. We can thus conclude that the nonequilibrium populations have a larger effect on the apparent MOKE signal than the relativistic spin-photon interaction.

 The demagnetization of Ni after an intensive laser pulse has recently been computed by Krieger \textit{et al.} \cite{krieger14}, who employed the time-dependent DFT formalism. Assuming extremely intense electromagnetic fields with a laser intensity of $10^{14} - 10^{15}$ W/m$^2$ they computed an appreciably larger demagnetization (of $\sim50$\,\%) than we do, which they attribute to dominance of \textit{nonlinear} effects. Conversely, in the current investigation we are in the moderate fluency regime, with typical laser intensities of $\simeq 10^{11}$ W/m$^2$, where the linear interaction Hamiltonian Eq.\ (\ref{eq:H_int}) should be sufficient and there is only a small number of electrons present in the excited state;  for this regime our results should be valid.

Summarizing, we performed the Foldy-Wouthuysen transformation on the Dirac-Kohn-Sham equation
in the presence of the exchange magnetic field as it is required for the relativistic density functional theory in the framework of the local spin density approximation. We obtained a Hamiltonian where several terms are consistent with results derived previously in Ref.\ \cite{crepieux01}.  We further showed that the spin-polarized term in the current density operator is irrelevant for the calculation of the conductivity spectra.
We discussed the modification caused by the relativistic spin-photon terms to the linear-response theory for the conductivity, and showed that an identical linear-response expression can be obtained for the nonrelativistic, semirelativistic and fully relativistic interaction Hamiltonians.  We then calculated the influence of the relativistic correction terms to the magneto-optical Kerr spectra of nickel. In the moderate fluency regime, where the linear-response theory is expected to be valid we find that relativistic spin-photon interactions can give a small modification ($\le 0.1$ \%) of the off-diagonal optical conductivity and of the MOKE signal. 
Thus, our calculations confirm that relativistic spin-photon interactions do exist, as originally proposed in Ref.\ \onlinecite{bigot09}, but we do not find that these could provide a notable channel of laser-induced magnetization loss.

\begin{acknowledgments}

 We thank P.\ Maldonado and A.\ Aperis  for valuable discussions.
 This work has been supported by 
the European Community's Seventh
Framework Programme (FP7/2007-2013) under grant agreement No.\ 281043, {}FEMTOSPIN,
the Swedish Research Council, and the Swedish National  Infrastructure for Computing (SNIC).	 

\end{acknowledgments}

\appendix
\section{The Pauli Hamiltonian}
\label{sec:pauli}
In the presence of an external electromagnetic field (characterized by the  vector potential $\bm{A}(\bm{r},t)$), the fully relativistic Dirac Hamiltonian (first without exchange field) has the form
 \begin{equation}
H=c\bm{\alpha}\cdot\left(\bm{p}-e\bm{A}\right)+\left(\bm{\beta} - \mathbb{1}\right)mc^{2}+V .
\label{pe} 
\end{equation}
%This is the relativistic DKS Hamiltonian in the presence of an external field. 
In the nonrelativistic limit (which is obtained by performing the FW transformation), it exactly gives the Pauli Hamiltonian
\begin{equation}
H_{\rm P}=\frac{\left(\bm{p}-e\bm{A}\right)^{2}}{2m}+V-\frac{e\hbar}{2m}\bm{\sigma}\cdot\bm{B} ,
\end{equation}
where the external magnetic field is given by $\bm{B}=\bm{\nabla}\times\bm{A}$. If we choose for simplicity a gauge such that
\[\bm{A}=\frac{\bm{B}\times\bm{r}}{2} ,\]
 which fulfills the Coulomb gauge ($\bm{\nabla}\cdot\bm{A}=0$) for the uniform magnetic field.
% (no spatial variation). 
Then, with this Pauli Hamiltonian one can show how the different magnetic contributions arise. 
 %The Pauli spin matrices are related to the spin by $\bm{S}=\frac{\hbar}{2}\bm{\sigma}$. 
 Namely, the Pauli Hamiltonian can be rewritten as 
\begin{equation}
H_{\rm P}= \left(\frac{p^{2}}{2m}+V\right)- \mu_B\, \bm{B}\cdot\left(\bm{L}+g\bm{S}\right)+\frac{e^{2}}{8m}\left(\bm{B}\times\bm{r}\right)^{2} ,
\end{equation}
with 
%$\mu_B$ the Bohr magneton and 
$g$ the Land\'e g-factor, which is 2 for spin degrees of freedom.  
The first term obviously is the unperturbed Hamiltonian, the dominant perturbation is the paramagnetic contribution and the last term is the diamagnetic contribution \cite{blundell01}.
Note that the external magnetic field couples to both the spin and orbital angular momentum operators, as it should be.

%\section{Extended Pauli Hamiltonian}
For magnetic materials the Pauli exclusion principle gives in addition rise to magnetic exchange. To include magnetic exchange (which dominates over the dipole-dipole interaction) the DKS Hamiltonian has to be written as \cite{eschrig96}
\begin{equation}
H=c\, \bm{\alpha}\cdot\left(\bm{p}-e\bm{A}\right)+\left({\bm \beta} -\mathbb{1}\right)mc^{2}+V + \mu_B {\bm \beta} \, \bm{\Sigma} \cdot\bm{B}^{\rm xc} ,
\label{xc}
\end{equation}
where the exchange field $\bm{B}^{\rm xc}$ has to be separated from the external magnetic vector potential as it would otherwise couple to the orbital degrees of freedom. The FW transformation of this Hamiltonian leads to the Hamiltonian given in Eq.\ (\ref{eq:FW_hamiltonian}). 
%Note that the Hamiltonian (\ref{eq:FW_hamiltonian}) is 
%different from the Hamiltonian given by Bigot \textit{et al.} \cite{bigot09}, as  they did not consider the magnetic exchange interaction, which however is the strongest interaction for a magnetic material (as, e.g., Fe, Co, or Ni).

\section{Derivation of the optical conductivity}
\label{sec:lin_resp}

Introducing the electromagnetic field $\bm{A}(\bm{r},t)$ produced by the intensive laser pulse the first-order interaction Hamiltonian could be written in terms of the momentum operator [Eq.\ (\ref{eq:rel_mom})] of the unperturbed Hamiltonian, 
%\newline
%The change in the Hamiltonian for charges distributed in space  when a magnetic vector potential is varied 
%is given by \cite{landau81}
\begin{equation}
 H_I=-\frac{e}{m}\bm{\Pi}\cdot\bm{A} .
\label{C1}
\end{equation}
%This momentum operator is proportional to the current density. 
This form of the interaction enters  in the nonrelativistic, semirelativistic
and fully relativistic cases. Here we consider the semirelativistic case corresponding to the extended Pauli Hamiltonian. We also note that, using a proper gauge, $\bm{A}=\frac{\bm{B}\times\bm{r}}{2}$, the above mentioned Hamiltonian can be rewritten as first-order interaction Hamiltonian in $\bm{E}(\bm{r},t)$. Using the gauge, it is obvious that $\bm{r}\cdot\bm{A}=0$, which gives
\begin{eqnarray}
\frac{d}{dt}(\bm{r}\cdot\bm{A}) &=& 0  \Rightarrow \dot{\bm r}\cdot\bm{A} =\bm{r}\cdot\bm{E}\nonumber\\
&& \Rightarrow\frac{e}{m}
\bm{\Pi}\cdot\bm{A} = e\bm{r}\cdot\bm{E} .
\end{eqnarray}
Therefore, the first-order interaction Hamiltonian can equally well be written in the form:
\begin{eqnarray}
H_I=-e \sum_i\bm{r}_i\cdot\bm{E}\equiv Be^{-i\omega^+ t} ,
\label{int_hamil}
\end{eqnarray}
where $\bm{r}_i$ are the positions of the electrons and $\omega^+\equiv\omega+i/\tau$.
In linear-response theory the total average, induced current $\bm{J}=\bm{j}V$, with $V$ the volume of the system, is computed from (see, e.g.\ \cite{callaway74}):
\begin{equation}
J (t)= \textrm{Tr}(\rho _0 J)+\frac{1}{i\hbar}\int_{-\infty}^{t} \! dt^{\prime}\big\langle\left[J(t),H_{I}(t^{\prime})\right]\big\rangle_{0},
\label{C2}
\end{equation}
where $\langle...\rangle_{0}$ means that the average has to be computed
with the equilibrium density matrix $\rho_{0}$. The first term refers to the equilibrium current density, which is usually taken to be zero in linear-response theory. Note  however the difference to the derivation given in Refs.\ \cite{callaway74,landau81},  where this  is not done and a second-order interaction term $\bm{A}^2$ is introduced in the Hamiltonian \cite{landau81}.  This term is rewritten in Refs.\ \cite{callaway74,landau81} and leads to the Drude response  (first term in (\ref{B6}) below). Such term should however not be included in a linear-response treatment, and it is actually not needed, as our derivation shows.
%The equilibrium current is given by the 2nd order interaction term $\bm{A}^2$ in the Hamiltonian \cite{landau81}, which is not in our case. 
Our formalism is valid in nonrelativistic, semirelativistic and fully relativistic case. 
%\begin{eqnarray}
%\bm{j}=\frac{ie\hbar}{2m}\left[\psi\bm{\nabla}\psi^{*}-\psi^*\bm{\nabla}\psi\right]-\frac{e^2}{m}\bm{A}\psi^*\psi,
%\end{eqnarray}
%where the Zeeman current has been dropped out. Notice that only the second term has an average  non-zero contribution in equilibrium. 
We introduce the linear interaction Hamiltonian according to (\ref{int_hamil}) in the second term of Eq.\ (\ref{C2}) and calculate the integral.  Partial integration of this second term leads to:
\begin{eqnarray}
J_{\alpha}(t) &=& \frac{1}{i\hbar}\big\langle\left[J_{\alpha}(t),B(t)\right]\big\rangle_{0}\frac{e^{-i\omega^+ t}}{-i\omega^+}\nonumber\\
&-&\frac{1}{i\hbar}\int_{-\infty}^t dt^{\prime}\big\langle [J_{\alpha}(t),\dot{B}(t^{\prime}) ]\big\rangle_{0}\frac{e^{-i\omega^+ t^{\prime}}}{-i\omega^+} ,
\end{eqnarray}
where
$\dot{B}$ is the derivative of $B$, which is related to the current, $\dot{B}(t)=-e\sum_i\dot{\bm{r}_i}(t)\cdot\bm{E}=-\bm{J}(t)\cdot\bm{E}$. 
Using Eq.\ (\ref{int_hamil}) we calculate the commutators and
it is evident that the integral leads to the current-current correlation in the average current, 
%Furthermore, we observe that the time derivative of the vector potential is the electric field,
%$\bm{E}=-\frac{d\bm{A}}{dt}$.
%The employed external field is oscillatory in nature, i.e.\  $E\propto e^{-i\omega^{+} t}$, thus $A\propto\frac{e^{-i\omega^{+} t}}{i\omega^{+}}=\frac{E}{i\omega^{+}}$. Realizing that, $\textrm{Tr}\left(\rho_{0}\psi^{*}\psi\right)=N$, the particle number, the average current density  is computed as 
\begin{eqnarray}
\!\!\! \!\!\! \!\!\! J_{\alpha}(t)&=&\frac{iNe^{2}}{m\omega^+}E_{\alpha}(t)e^{-i\omega^+ t}\nonumber\\&-&\frac{1}{i\hbar}\int_{-\infty}^{t} \!\!\! dt^{\prime}\big\langle\left[ J_{\alpha}(t),J_{\beta} (t^{\prime}) \right]\big\rangle_{0}\frac{E_{\beta}(t^{\prime})e^{-i\omega^+ t^{\prime}}}{i\omega^{+}} .
\label{B6}
\end{eqnarray}
The conductivity response to the electromagnetic field is given as
\begin{equation}
j_{\alpha}(t)=\int_{-\infty}^{t}\! dt^{\prime}\sigma_{\alpha\beta} (t-t^{\prime})E_{\beta}(t^{\prime}) .
\label{4}
\end{equation}

Now, comparing both equations we obtain the linear-response expression for the conductivity. Computed in Fourier space, the conductivity in terms of (noninteracting) single-particle states is then (see Ref.\ \cite{oppeneer01} for details)
\begin{eqnarray}
\sigma_{\alpha\beta}(\omega) &=& \frac{ie^2N\delta_{\alpha\beta}}{Vm\omega^+}\nonumber\\
&+&\frac{i}{V\hbar \omega^+}\sum_{nn'} \frac{f(\epsilon_n)-f(\epsilon_{n'})}{{\omega^+-\omega_{n^{\prime}n}}}j^{\alpha}_{nn^{\prime}}j^{\beta}_{n^{\prime}n},
\label{conductivity}
\end{eqnarray}
where $j_{nn'}$ are the matrix elements of the current density operator for the single-particle states $n$ and $n'$, $f(\epsilon_n)$ is the Fermi-Dirac distribution function of the $n$-th state having energy $\epsilon_n$ and $\hbar\omega_{n'n}=\epsilon_{n'}-\epsilon_{n}$. This  linear-response expression is exact for the non-relativistic, semirelativistic and fully relativistic cases. In the semirelativistic limit, the two terms in Eq.\ (\ref{conductivity}) can be approximately joined together, which yields
\begin{equation}
\sigma_{\alpha\beta}(\omega)\approx-\frac{i}{\hbar V}\sum_{nn'} \frac{f(\epsilon_n)-f(\epsilon_{n'})}{\omega_{nn'}}\frac{j^{\alpha}_{n'n}j^{\beta}_{nn'}}{\omega-\omega_{nn'}+{i}/{\tau}},
\end{equation}
 where the $\approx $ sign relates to the intraband term {i.e., $n=n'$) for which the approximation $\bm{\Pi}_{\rm P} \approx \bm{p}$ has been made.\\

\section{Derivation of the spin-polarized current}
\label{zeeman_current}

 The current density operator in semirelativistic form was previously shown \cite{nowakowski99,huszar67} to contain a term $\bm{j}_{sp}=  -\frac{ie}{2m}(\overleftarrow{\bm{p}}\times \bm{\sigma}+ \bm{\sigma}\times{\bm{p}})$. This term also appears in our treatment.
When we apply the FW transformation to the DKS Hamiltonian in Eq.\ (\ref{eq:DKSH_xc}), we find a term $\frac{i}{2m}\bm{\sigma}\cdot(\bm{p}\times\bm{p})$, in the Hamiltonian. This term is taken to be zero for obvious reasons. However, it is this term that leads to the spin-polarized current density $\bm{j}_{sp}$.

Defining the charge density as $\rho=e\delta\left(\bm{r}-\hat{\bm{r}}\right)$ and 
using the Heisenberg equation of motion for the above-given Hamiltonian term,
\begin{eqnarray}
\frac{d\rho}{dt} &=& \frac{1}{i\hbar}\left[e\delta\left(\bm{r}-\hat{\bm{r}}\right),\frac{i}{2m}\bm{\sigma}\cdot(\bm{p}\times\bm{p})\right]\nonumber\\
%&=& \frac{e}{2m\hbar}\bm{\sigma}\cdot\left[\delta\left(\bm{r}-\hat{\bm{r}}\right),(\bm{p}\times\bm{p})\right]\nonumber\\
&=& \frac{e}{2m\hbar}\bm{\sigma}\cdot\left\{\left[\delta\left(\bm{r}-\hat{\bm{r}}\right),\bm{p}\right]\times\bm{p}+\bm{p}\times\left[\delta\left(\bm{r}-\hat{\bm{r}}\right),\bm{p}\right]\right\}\nonumber\\
%&=& \frac{ie}{2m}\bm{\sigma}\cdot\left\{|\bm{r}\rangle\nabla_{\bm{r}}\langle\bm{r}|
%\times\bm{p}+\bm{p}\times \bm{r}\rangle\nabla_{\bm{r}}\langle\bm{r}|\right\}\nonumber\\
%&=& \frac{ie}{2m}\bm{\nabla}_{\bm{r}}\left\{\vert\bm{r}\rangle\bm{\sigma}\langle\bm{r}\vert
%\times\bm{p}+\bm{p}\times \vert\bm{r}\rangle\bm{\sigma}\langle\bm{r}\vert\right\}\nonumber\\ 
&=& \bm{\nabla}\cdot\frac{ie}{2m}\left\{\bm{\bm{\sigma}}
\times\bm{p}+\overleftarrow{\bm{p}}\times\bm{\bm{\sigma}}\right\}\nonumber\\
&=& -\bm{\nabla}\cdot\bm{j}_{sp} .
\end{eqnarray}
In this derivation, we make use of the fact that, the commutator in the position basis, $\left[\delta\left(\bm{r}-\hat{\bm{r}}\right),\bm{p}\right]=i\hbar\vert \bm{r}\rangle\bm{\nabla}_{\bm{r}}\langle\bm{r}\vert$. Using the continuity equation we extract the spin-polarized current density operator as
 %\begin{eqnarray}
${\bm{j}}_{sp}=-\frac{ie}{2m}  \{ \bm{\sigma}
\times \bm{p}+ \overleftarrow{\bm{p}}\times {\bm{\sigma}} \}  $.
%\end{eqnarray}
%
The matrix elements are given by:
\begin{eqnarray}
\!\! \langle\Psi_n\vert\bm{j}_{sp}\vert\Psi_m\rangle &=& -\frac{e\hbar}{2m}\int_V \!\! d\bm{r}\left\{\Psi_n^*\bm{\sigma}\times\bm{\nabla}\Psi_m \! - \! \Psi_m\bm{\nabla}\Psi_n^*\times\bm{\sigma}\right\}\nonumber\\
&=& \frac{e\hbar}{2m}\int_V \!\! d\bm{r}\bm{\nabla}\times\left[\Psi_n^*\bm{\sigma}\Psi_m\right].
\end{eqnarray}
%{\blue In Maxwell theory this term corresponds to a current due to the light-induced magnetization variation, $\propto \bm{\nabla} \times \bm{M}$, which can be neglected at optical frequencies \cite{landau84}.}
%\bibliographystyle{unsrt}
%\bibliographystyle{apsrev4-1}
%\bibliography{biblio}

\begin{thebibliography}{62}%
\makeatletter
\providecommand \@ifxundefined [1]{%
 \@ifx{#1\undefined}
}%
\providecommand \@ifnum [1]{%
 \ifnum #1\expandafter \@firstoftwo
 \else \expandafter \@secondoftwo
 \fi
}%
\providecommand \@ifx [1]{%
 \ifx #1\expandafter \@firstoftwo
 \else \expandafter \@secondoftwo
 \fi
}%
\providecommand \natexlab [1]{#1}%
\providecommand \enquote  [1]{``#1''}%
\providecommand \bibnamefont  [1]{#1}%
\providecommand \bibfnamefont [1]{#1}%
\providecommand \citenamefont [1]{#1}%
\providecommand \href@noop [0]{\@secondoftwo}%
\providecommand \href [0]{\begingroup \@sanitize@url \@href}%
\providecommand \@href[1]{\@@startlink{#1}\@@href}%
\providecommand \@@href[1]{\endgroup#1\@@endlink}%
\providecommand \@sanitize@url [0]{\catcode `\\12\catcode `\$12\catcode
  `\&12\catcode `\#12\catcode `\^12\catcode `\_12\catcode `\%12\relax}%
\providecommand \@@startlink[1]{}%
\providecommand \@@endlink[0]{}%
\providecommand \url  [0]{\begingroup\@sanitize@url \@url }%
\providecommand \@url [1]{\endgroup\@href {#1}{\urlprefix }}%
\providecommand \urlprefix  [0]{URL }%
\providecommand \Eprint [0]{\href }%
\providecommand \doibase [0]{http://dx.doi.org/}%
\providecommand \selectlanguage [0]{\@gobble}%
\providecommand \bibinfo  [0]{\@secondoftwo}%
\providecommand \bibfield  [0]{\@secondoftwo}%
\providecommand \translation [1]{[#1]}%
\providecommand \BibitemOpen [0]{}%
\providecommand \bibitemStop [0]{}%
\providecommand \bibitemNoStop [0]{.\EOS\space}%
\providecommand \EOS [0]{\spacefactor3000\relax}%
\providecommand \BibitemShut  [1]{\csname bibitem#1\endcsname}%
\let\auto@bib@innerbib\@empty
%</preamble>
\bibitem [{\citenamefont {Beaurepaire}\ \emph {et~al.}(1996)\citenamefont
  {Beaurepaire}, \citenamefont {Merle}, \citenamefont {Daunois},\ and\
  \citenamefont {Bigot}}]{beaurepaire96}%
  \BibitemOpen
  \bibfield  {author} {\bibinfo {author} {\bibfnamefont {E.}~\bibnamefont
  {Beaurepaire}}, \bibinfo {author} {\bibfnamefont {J.-C.}\ \bibnamefont
  {Merle}}, \bibinfo {author} {\bibfnamefont {A.}~\bibnamefont {Daunois}}, \
  and\ \bibinfo {author} {\bibfnamefont {J.-Y.}\ \bibnamefont {Bigot}},\ }\href
  {\doibase 10.1103/PhysRevLett.76.4250} {\bibfield  {journal} {\bibinfo
  {journal} {Phys. Rev. Lett.}\ }\textbf {\bibinfo {volume} {76}},\ \bibinfo
  {pages} {4250} (\bibinfo {year} {1996})}\BibitemShut {NoStop}%
\bibitem [{\citenamefont {Hohlfeld}\ \emph {et~al.}(1997)\citenamefont
  {Hohlfeld}, \citenamefont {Matthias}, \citenamefont {Knorren},\ and\
  \citenamefont {Bennemann}}]{hohlfeld97}%
  \BibitemOpen
  \bibfield  {author} {\bibinfo {author} {\bibfnamefont {J.}~\bibnamefont
  {Hohlfeld}}, \bibinfo {author} {\bibfnamefont {E.}~\bibnamefont {Matthias}},
  \bibinfo {author} {\bibfnamefont {R.}~\bibnamefont {Knorren}}, \ and\
  \bibinfo {author} {\bibfnamefont {K.~H.}\ \bibnamefont {Bennemann}},\ }\href
  {\doibase 10.1103/PhysRevLett.78.4861} {\bibfield  {journal} {\bibinfo
  {journal} {Phys. Rev. Lett.}\ }\textbf {\bibinfo {volume} {78}},\ \bibinfo
  {pages} {4861} (\bibinfo {year} {1997})}\BibitemShut {NoStop}%
\bibitem [{\citenamefont {Scholl}\ \emph {et~al.}(1997)\citenamefont {Scholl},
  \citenamefont {Baumgarten}, \citenamefont {Jacquemin},\ and\ \citenamefont
  {Eberhardt}}]{scholl97}%
  \BibitemOpen
  \bibfield  {author} {\bibinfo {author} {\bibfnamefont {A.}~\bibnamefont
  {Scholl}}, \bibinfo {author} {\bibfnamefont {L.}~\bibnamefont {Baumgarten}},
  \bibinfo {author} {\bibfnamefont {R.}~\bibnamefont {Jacquemin}}, \ and\
  \bibinfo {author} {\bibfnamefont {W.}~\bibnamefont {Eberhardt}},\ }\href
  {\doibase 10.1103/PhysRevLett.79.5146} {\bibfield  {journal} {\bibinfo
  {journal} {Phys. Rev. Lett.}\ }\textbf {\bibinfo {volume} {79}},\ \bibinfo
  {pages} {5146} (\bibinfo {year} {1997})}\BibitemShut {NoStop}%
\bibitem [{\citenamefont {Regensburger}\ \emph {et~al.}(2000)\citenamefont
  {Regensburger}, \citenamefont {Vollmer},\ and\ \citenamefont
  {Kirschner}}]{regensburger00}%
  \BibitemOpen
  \bibfield  {author} {\bibinfo {author} {\bibfnamefont {H.}~\bibnamefont
  {Regensburger}}, \bibinfo {author} {\bibfnamefont {R.}~\bibnamefont
  {Vollmer}}, \ and\ \bibinfo {author} {\bibfnamefont {J.}~\bibnamefont
  {Kirschner}},\ }\href {\doibase 10.1103/PhysRevB.61.14716} {\bibfield
  {journal} {\bibinfo  {journal} {Phys. Rev. B}\ }\textbf {\bibinfo {volume}
  {61}},\ \bibinfo {pages} {14716} (\bibinfo {year} {2000})}\BibitemShut
  {NoStop}%
\bibitem [{\citenamefont {Kampfrath}\ \emph {et~al.}(2002)\citenamefont
  {Kampfrath}, \citenamefont {Ulbrich}, \citenamefont {Leuenberger},
  \citenamefont {M{\"u}nzenberg}, \citenamefont {Sass},\ and\ \citenamefont
  {Felsch}}]{kampfrath02}%
  \BibitemOpen
  \bibfield  {author} {\bibinfo {author} {\bibfnamefont {T.}~\bibnamefont
  {Kampfrath}}, \bibinfo {author} {\bibfnamefont {R.~G.}\ \bibnamefont
  {Ulbrich}}, \bibinfo {author} {\bibfnamefont {F.}~\bibnamefont
  {Leuenberger}}, \bibinfo {author} {\bibfnamefont {M.}~\bibnamefont
  {M{\"u}nzenberg}}, \bibinfo {author} {\bibfnamefont {B.}~\bibnamefont
  {Sass}}, \ and\ \bibinfo {author} {\bibfnamefont {W.}~\bibnamefont
  {Felsch}},\ }\href@noop {} {\bibfield  {journal} {\bibinfo  {journal} {Phys.
  Rev. B}\ }\textbf {\bibinfo {volume} {65}},\ \bibinfo {pages} {104429}
  (\bibinfo {year} {2002})}\BibitemShut {NoStop}%
\bibitem [{\citenamefont {van Kampen}\ \emph {et~al.}(2005)\citenamefont {van
  Kampen}, \citenamefont {Kohlhepp}, \citenamefont {de~Jonge}, \citenamefont
  {Koopmans},\ and\ \citenamefont {Coehoorn}}]{vankampen05}%
  \BibitemOpen
  \bibfield  {author} {\bibinfo {author} {\bibfnamefont {M.}~\bibnamefont {van
  Kampen}}, \bibinfo {author} {\bibfnamefont {J.~T.}\ \bibnamefont {Kohlhepp}},
  \bibinfo {author} {\bibfnamefont {W.~J.~M.}\ \bibnamefont {de~Jonge}},
  \bibinfo {author} {\bibfnamefont {B.}~\bibnamefont {Koopmans}}, \ and\
  \bibinfo {author} {\bibfnamefont {R.}~\bibnamefont {Coehoorn}},\ }\href@noop
  {} {\bibfield  {journal} {\bibinfo  {journal} {J. Phys.: Condens. Matter}\
  }\textbf {\bibinfo {volume} {17}},\ \bibinfo {pages} {6823} (\bibinfo {year}
  {2005})}\BibitemShut {NoStop}%
\bibitem [{\citenamefont {Cheskis}\ \emph {et~al.}(2005)\citenamefont
  {Cheskis}, \citenamefont {Porat}, \citenamefont {Szapiro}, \citenamefont
  {Potashnik},\ and\ \citenamefont {Bar-Ad}}]{cheskis05}%
  \BibitemOpen
  \bibfield  {author} {\bibinfo {author} {\bibfnamefont {D.}~\bibnamefont
  {Cheskis}}, \bibinfo {author} {\bibfnamefont {A.}~\bibnamefont {Porat}},
  \bibinfo {author} {\bibfnamefont {L.}~\bibnamefont {Szapiro}}, \bibinfo
  {author} {\bibfnamefont {O.}~\bibnamefont {Potashnik}}, \ and\ \bibinfo
  {author} {\bibfnamefont {S.}~\bibnamefont {Bar-Ad}},\ }\href@noop {}
  {\bibfield  {journal} {\bibinfo  {journal} {Phys. Rev. B}\ }\textbf {\bibinfo
  {volume} {72}},\ \bibinfo {pages} {014437} (\bibinfo {year}
  {2005})}\BibitemShut {NoStop}%
\bibitem [{\citenamefont {Carley}\ \emph {et~al.}(2012)\citenamefont {Carley},
  \citenamefont {D\"obrich}, \citenamefont {Frietsch}, \citenamefont {Gahl},
  \citenamefont {Teichmann}, \citenamefont {Schwarzkopf}, \citenamefont
  {Wernet},\ and\ \citenamefont {Weinelt}}]{carley12}%
  \BibitemOpen
  \bibfield  {author} {\bibinfo {author} {\bibfnamefont {R.}~\bibnamefont
  {Carley}}, \bibinfo {author} {\bibfnamefont {K.}~\bibnamefont {D\"obrich}},
  \bibinfo {author} {\bibfnamefont {B.}~\bibnamefont {Frietsch}}, \bibinfo
  {author} {\bibfnamefont {C.}~\bibnamefont {Gahl}}, \bibinfo {author}
  {\bibfnamefont {M.}~\bibnamefont {Teichmann}}, \bibinfo {author}
  {\bibfnamefont {O.}~\bibnamefont {Schwarzkopf}}, \bibinfo {author}
  {\bibfnamefont {P.}~\bibnamefont {Wernet}}, \ and\ \bibinfo {author}
  {\bibfnamefont {M.}~\bibnamefont {Weinelt}},\ }\href {\doibase
  10.1103/PhysRevLett.109.057401} {\bibfield  {journal} {\bibinfo  {journal}
  {Phys. Rev. Lett.}\ }\textbf {\bibinfo {volume} {109}},\ \bibinfo {pages}
  {0574012} (\bibinfo {year} {2012})}\BibitemShut {NoStop}%
\bibitem [{\citenamefont {Boeglin}\ \emph {et~al.}(2010)\citenamefont
  {Boeglin}, \citenamefont {Beaurepaire}, \citenamefont {Halt{\'e}},
  \citenamefont {L{\'o}pez-Flores}, \citenamefont {Stamm}, \citenamefont
  {Pontius}, \citenamefont {D{\"u}rr},\ and\ \citenamefont
  {Bigot}}]{boeglin10}%
  \BibitemOpen
  \bibfield  {author} {\bibinfo {author} {\bibfnamefont {C.}~\bibnamefont
  {Boeglin}}, \bibinfo {author} {\bibfnamefont {E.}~\bibnamefont
  {Beaurepaire}}, \bibinfo {author} {\bibfnamefont {V.}~\bibnamefont
  {Halt{\'e}}}, \bibinfo {author} {\bibfnamefont {V.}~\bibnamefont
  {L{\'o}pez-Flores}}, \bibinfo {author} {\bibfnamefont {C.}~\bibnamefont
  {Stamm}}, \bibinfo {author} {\bibfnamefont {N.}~\bibnamefont {Pontius}},
  \bibinfo {author} {\bibfnamefont {H.~A.}\ \bibnamefont {D{\"u}rr}}, \ and\
  \bibinfo {author} {\bibfnamefont {J.-Y.}\ \bibnamefont {Bigot}},\ }\href
  {\doibase doi:10.1038/nature09070} {\bibfield  {journal} {\bibinfo  {journal}
  {Nature}\ }\textbf {\bibinfo {volume} {465}},\ \bibinfo {pages} {458}
  (\bibinfo {year} {2010})}\BibitemShut {NoStop}%
\bibitem [{\citenamefont {Radu}\ \emph {et~al.}(2011)\citenamefont {Radu},
  \citenamefont {Vahaplar}, \citenamefont {Stamm}, \citenamefont {Kachel},
  \citenamefont {Pontius}, \citenamefont {Duerr}, \citenamefont {Ostler},
  \citenamefont {Barker}, \citenamefont {Evans}, \citenamefont {Chantrell},
  \citenamefont {Tsukamoto}, \citenamefont {Itoh}, \citenamefont {Kirilyuk},
  \citenamefont {Rasing},\ and\ \citenamefont {Kimel}}]{radu11}%
  \BibitemOpen
  \bibfield  {author} {\bibinfo {author} {\bibfnamefont {I.}~\bibnamefont
  {Radu}}, \bibinfo {author} {\bibfnamefont {K.}~\bibnamefont {Vahaplar}},
  \bibinfo {author} {\bibfnamefont {C.}~\bibnamefont {Stamm}}, \bibinfo
  {author} {\bibfnamefont {T.}~\bibnamefont {Kachel}}, \bibinfo {author}
  {\bibfnamefont {N.}~\bibnamefont {Pontius}}, \bibinfo {author} {\bibfnamefont
  {H.~A.}\ \bibnamefont {Duerr}}, \bibinfo {author} {\bibfnamefont {T.~A.}\
  \bibnamefont {Ostler}}, \bibinfo {author} {\bibfnamefont {J.}~\bibnamefont
  {Barker}}, \bibinfo {author} {\bibfnamefont {R.~F.~L.}\ \bibnamefont
  {Evans}}, \bibinfo {author} {\bibfnamefont {R.~W.}\ \bibnamefont
  {Chantrell}}, \bibinfo {author} {\bibfnamefont {A.}~\bibnamefont
  {Tsukamoto}}, \bibinfo {author} {\bibfnamefont {A.}~\bibnamefont {Itoh}},
  \bibinfo {author} {\bibfnamefont {A.}~\bibnamefont {Kirilyuk}}, \bibinfo
  {author} {\bibfnamefont {{\rm Th}.}~\bibnamefont {Rasing}}, \ and\ \bibinfo
  {author} {\bibfnamefont {A.~V.}\ \bibnamefont {Kimel}},\ }\href@noop {}
  {\bibfield  {journal} {\bibinfo  {journal} {Nature}\ }\textbf {\bibinfo
  {volume} {472}},\ \bibinfo {pages} {205} (\bibinfo {year}
  {2011})}\BibitemShut {NoStop}%
\bibitem [{\citenamefont {Mathias}\ \emph {et~al.}(2012)\citenamefont
  {Mathias}, \citenamefont {La-O-Vorakiat}, \citenamefont {Grychtol},
  \citenamefont {Granitzka}, \citenamefont {Turgut}, \citenamefont {Shaw},
  \citenamefont {Adam}, \citenamefont {Nembach}, \citenamefont {Siemens},
  \citenamefont {Eich}, \citenamefont {Schneider}, \citenamefont {Silva},
  \citenamefont {Aeschlimann}, \citenamefont {Murnane},\ and\ \citenamefont
  {Kapteyn}}]{mathias12}%
  \BibitemOpen
  \bibfield  {author} {\bibinfo {author} {\bibfnamefont {S.}~\bibnamefont
  {Mathias}}, \bibinfo {author} {\bibfnamefont {C.}~\bibnamefont
  {La-O-Vorakiat}}, \bibinfo {author} {\bibfnamefont {P.}~\bibnamefont
  {Grychtol}}, \bibinfo {author} {\bibfnamefont {P.}~\bibnamefont {Granitzka}},
  \bibinfo {author} {\bibfnamefont {E.}~\bibnamefont {Turgut}}, \bibinfo
  {author} {\bibfnamefont {J.~M.}\ \bibnamefont {Shaw}}, \bibinfo {author}
  {\bibfnamefont {R.}~\bibnamefont {Adam}}, \bibinfo {author} {\bibfnamefont
  {H.~T.}\ \bibnamefont {Nembach}}, \bibinfo {author} {\bibfnamefont {M.~E.}\
  \bibnamefont {Siemens}}, \bibinfo {author} {\bibfnamefont {S.}~\bibnamefont
  {Eich}}, \bibinfo {author} {\bibfnamefont {C.~M.}\ \bibnamefont {Schneider}},
  \bibinfo {author} {\bibfnamefont {T.~J.}\ \bibnamefont {Silva}}, \bibinfo
  {author} {\bibfnamefont {M.}~\bibnamefont {Aeschlimann}}, \bibinfo {author}
  {\bibfnamefont {M.~M.}\ \bibnamefont {Murnane}}, \ and\ \bibinfo {author}
  {\bibfnamefont {H.~C.}\ \bibnamefont {Kapteyn}},\ }\href@noop {} {\bibfield
  {journal} {\bibinfo  {journal} {Proc. Natl. Acad. Scie. USA}\ }\textbf
  {\bibinfo {volume} {109}},\ \bibinfo {pages} {4792} (\bibinfo {year}
  {2012})}\BibitemShut {NoStop}%
\bibitem [{\citenamefont {Rudolf}\ \emph {et~al.}(2012)\citenamefont {Rudolf},
  \citenamefont {La-O-Vorakiat}, \citenamefont {Battiato}, \citenamefont
  {Adam}, \citenamefont {Shaw}, \citenamefont {Turgut}, \citenamefont
  {Maldonado}, \citenamefont {Mathias}, \citenamefont {Grychtol}, \citenamefont
  {Nembach}, \citenamefont {Silva}, \citenamefont {Aeschlimann}, \citenamefont
  {Kapteyn}, \citenamefont {Murnane}, \citenamefont {Schneider},\ and\
  \citenamefont {Oppeneer}}]{rudolf12}%
  \BibitemOpen
  \bibfield  {author} {\bibinfo {author} {\bibfnamefont {D.}~\bibnamefont
  {Rudolf}}, \bibinfo {author} {\bibfnamefont {C.}~\bibnamefont
  {La-O-Vorakiat}}, \bibinfo {author} {\bibfnamefont {M.}~\bibnamefont
  {Battiato}}, \bibinfo {author} {\bibfnamefont {R.}~\bibnamefont {Adam}},
  \bibinfo {author} {\bibfnamefont {J.~M.}\ \bibnamefont {Shaw}}, \bibinfo
  {author} {\bibfnamefont {E.}~\bibnamefont {Turgut}}, \bibinfo {author}
  {\bibfnamefont {P.}~\bibnamefont {Maldonado}}, \bibinfo {author}
  {\bibfnamefont {S.}~\bibnamefont {Mathias}}, \bibinfo {author} {\bibfnamefont
  {P.}~\bibnamefont {Grychtol}}, \bibinfo {author} {\bibfnamefont {H.~T.}\
  \bibnamefont {Nembach}}, \bibinfo {author} {\bibfnamefont {T.~J.}\
  \bibnamefont {Silva}}, \bibinfo {author} {\bibfnamefont {M.}~\bibnamefont
  {Aeschlimann}}, \bibinfo {author} {\bibfnamefont {H.~C.}\ \bibnamefont
  {Kapteyn}}, \bibinfo {author} {\bibfnamefont {M.~M.}\ \bibnamefont
  {Murnane}}, \bibinfo {author} {\bibfnamefont {C.~M.}\ \bibnamefont
  {Schneider}}, \ and\ \bibinfo {author} {\bibfnamefont {P.~M.}\ \bibnamefont
  {Oppeneer}},\ }\href@noop {} {\bibfield  {journal} {\bibinfo  {journal}
  {Nature Commun.}\ }\textbf {\bibinfo {volume} {3}},\ \bibinfo {pages} {1037}
  (\bibinfo {year} {2012})}\BibitemShut {NoStop}%
\bibitem [{\citenamefont {Eschenlohr}\ \emph {et~al.}(2013)\citenamefont
  {Eschenlohr}, \citenamefont {Battiato}, \citenamefont {Maldonado},
  \citenamefont {Pontius}, \citenamefont {Kachel}, \citenamefont {Holldack},
  \citenamefont {Mitzner}, \citenamefont {F\"ohlisch}, \citenamefont
  {Oppeneer},\ and\ \citenamefont {Stamm}}]{eschenlohr13}%
  \BibitemOpen
  \bibfield  {author} {\bibinfo {author} {\bibfnamefont {A.}~\bibnamefont
  {Eschenlohr}}, \bibinfo {author} {\bibfnamefont {M.}~\bibnamefont
  {Battiato}}, \bibinfo {author} {\bibfnamefont {P.}~\bibnamefont {Maldonado}},
  \bibinfo {author} {\bibfnamefont {N.}~\bibnamefont {Pontius}}, \bibinfo
  {author} {\bibfnamefont {T.}~\bibnamefont {Kachel}}, \bibinfo {author}
  {\bibfnamefont {K.}~\bibnamefont {Holldack}}, \bibinfo {author}
  {\bibfnamefont {R.}~\bibnamefont {Mitzner}}, \bibinfo {author} {\bibfnamefont
  {A.}~\bibnamefont {F\"ohlisch}}, \bibinfo {author} {\bibfnamefont {P.~M.}\
  \bibnamefont {Oppeneer}}, \ and\ \bibinfo {author} {\bibfnamefont
  {C.}~\bibnamefont {Stamm}},\ }\href@noop {} {\bibfield  {journal} {\bibinfo
  {journal} {Nature Mater.}\ }\textbf {\bibinfo {volume} {12}},\ \bibinfo
  {pages} {332} (\bibinfo {year} {2013})}\BibitemShut {NoStop}%
\bibitem [{\citenamefont {Bergeard}\ \emph {et~al.}(2014)\citenamefont
  {Bergeard}, \citenamefont {L{\'o}pez-Flores}, \citenamefont {Halt{\'e}},
  \citenamefont {Hehn}, \citenamefont {Stamm}, \citenamefont {Pontius},
  \citenamefont {Beaurepaire},\ and\ \citenamefont {Boeglin}}]{bergeard14}%
  \BibitemOpen
  \bibfield  {author} {\bibinfo {author} {\bibfnamefont {N.}~\bibnamefont
  {Bergeard}}, \bibinfo {author} {\bibfnamefont {V.}~\bibnamefont
  {L{\'o}pez-Flores}}, \bibinfo {author} {\bibfnamefont {V.}~\bibnamefont
  {Halt{\'e}}}, \bibinfo {author} {\bibfnamefont {M.}~\bibnamefont {Hehn}},
  \bibinfo {author} {\bibfnamefont {C.}~\bibnamefont {Stamm}}, \bibinfo
  {author} {\bibfnamefont {N.}~\bibnamefont {Pontius}}, \bibinfo {author}
  {\bibfnamefont {E.}~\bibnamefont {Beaurepaire}}, \ and\ \bibinfo {author}
  {\bibfnamefont {C.}~\bibnamefont {Boeglin}},\ }\href {\doibase
  doi:10.1038/ncomms4466} {\bibfield  {journal} {\bibinfo  {journal} {Nature
  Commun.}\ }\textbf {\bibinfo {volume} {5}},\ \bibinfo {pages} {3466}
  (\bibinfo {year} {2014})}\BibitemShut {NoStop}%
\bibitem [{\citenamefont {Bovensiepen}(2009)}]{bovensiepen09}%
  \BibitemOpen
  \bibfield  {author} {\bibinfo {author} {\bibfnamefont {U.}~\bibnamefont
  {Bovensiepen}},\ }\href@noop {} {\bibfield  {journal} {\bibinfo  {journal}
  {Nat. Phys.}\ }\textbf {\bibinfo {volume} {5}},\ \bibinfo {pages} {461}
  (\bibinfo {year} {2009})}\BibitemShut {NoStop}%
\bibitem [{\citenamefont {Kirilyuk}\ \emph {et~al.}(2010)\citenamefont
  {Kirilyuk}, \citenamefont {Kimel},\ and\ \citenamefont
  {Rasing}}]{kirilyuk10}%
  \BibitemOpen
  \bibfield  {author} {\bibinfo {author} {\bibfnamefont {A.}~\bibnamefont
  {Kirilyuk}}, \bibinfo {author} {\bibfnamefont {A.~V.}\ \bibnamefont {Kimel}},
  \ and\ \bibinfo {author} {\bibfnamefont {{\rm Th}.}~\bibnamefont {Rasing}},\
  }\href@noop {} {\bibfield  {journal} {\bibinfo  {journal} {Rev. Mod. Phys.}\
  }\textbf {\bibinfo {volume} {82}},\ \bibinfo {pages} {2731} (\bibinfo {year}
  {2010})}\BibitemShut {NoStop}%
\bibitem [{\citenamefont {Carva}\ \emph
  {et~al.}(2011{\natexlab{a}})\citenamefont {Carva}, \citenamefont {Battiato},\
  and\ \citenamefont {Oppeneer}}]{carva11nat}%
  \BibitemOpen
  \bibfield  {author} {\bibinfo {author} {\bibfnamefont {K.}~\bibnamefont
  {Carva}}, \bibinfo {author} {\bibfnamefont {M.}~\bibnamefont {Battiato}}, \
  and\ \bibinfo {author} {\bibfnamefont {P.~M.}\ \bibnamefont {Oppeneer}},\
  }\href@noop {} {\bibfield  {journal} {\bibinfo  {journal} {Nature Phys.}\
  }\textbf {\bibinfo {volume} {7}},\ \bibinfo {pages} {665} (\bibinfo {year}
  {2011}{\natexlab{a}})}\BibitemShut {NoStop}%
\bibitem [{\citenamefont {Zhang}\ and\ \citenamefont
  {H\"ubner}(2000)}]{zhang00}%
  \BibitemOpen
  \bibfield  {author} {\bibinfo {author} {\bibfnamefont {G.~P.}\ \bibnamefont
  {Zhang}}\ and\ \bibinfo {author} {\bibfnamefont {W.}~\bibnamefont
  {H\"ubner}},\ }\href {\doibase 10.1103/PhysRevLett.85.3025} {\bibfield
  {journal} {\bibinfo  {journal} {Phys. Rev. Lett.}\ }\textbf {\bibinfo
  {volume} {85}},\ \bibinfo {pages} {3025} (\bibinfo {year}
  {2000})}\BibitemShut {NoStop}%
\bibitem [{\citenamefont {Carpene}\ \emph {et~al.}(2008)\citenamefont
  {Carpene}, \citenamefont {Mancini}, \citenamefont {Dallera}, \citenamefont
  {Brenna}, \citenamefont {Puppin},\ and\ \citenamefont
  {De~Silvestri}}]{carpene08}%
  \BibitemOpen
  \bibfield  {author} {\bibinfo {author} {\bibfnamefont {E.}~\bibnamefont
  {Carpene}}, \bibinfo {author} {\bibfnamefont {E.}~\bibnamefont {Mancini}},
  \bibinfo {author} {\bibfnamefont {C.}~\bibnamefont {Dallera}}, \bibinfo
  {author} {\bibfnamefont {M.}~\bibnamefont {Brenna}}, \bibinfo {author}
  {\bibfnamefont {E.}~\bibnamefont {Puppin}}, \ and\ \bibinfo {author}
  {\bibfnamefont {S.}~\bibnamefont {De~Silvestri}},\ }\href@noop {} {\bibfield
  {journal} {\bibinfo  {journal} {Phys. Rev. B}\ }\textbf {\bibinfo {volume}
  {78}} (\bibinfo {year} {2008})}\BibitemShut {NoStop}%
\bibitem [{\citenamefont {Koopmans}\ \emph {et~al.}(2010)\citenamefont
  {Koopmans}, \citenamefont {Malinowski}, \citenamefont {Dalla~Longa},
  \citenamefont {Steiauf}, \citenamefont {Faehnle}, \citenamefont {Roth},
  \citenamefont {Cinchetti},\ and\ \citenamefont {Aeschlimann}}]{koopmans10}%
  \BibitemOpen
  \bibfield  {author} {\bibinfo {author} {\bibfnamefont {B.}~\bibnamefont
  {Koopmans}}, \bibinfo {author} {\bibfnamefont {G.}~\bibnamefont
  {Malinowski}}, \bibinfo {author} {\bibfnamefont {F.}~\bibnamefont
  {Dalla~Longa}}, \bibinfo {author} {\bibfnamefont {D.}~\bibnamefont
  {Steiauf}}, \bibinfo {author} {\bibfnamefont {M.}~\bibnamefont {Faehnle}},
  \bibinfo {author} {\bibfnamefont {T.}~\bibnamefont {Roth}}, \bibinfo {author}
  {\bibfnamefont {M.}~\bibnamefont {Cinchetti}}, \ and\ \bibinfo {author}
  {\bibfnamefont {M.}~\bibnamefont {Aeschlimann}},\ }\href@noop {} {\bibfield
  {journal} {\bibinfo  {journal} {Nature Mater.}\ }\textbf {\bibinfo {volume}
  {9}},\ \bibinfo {pages} {259} (\bibinfo {year} {2010})}\BibitemShut {NoStop}%
\bibitem [{\citenamefont {Krauss}\ \emph {et~al.}(2009)\citenamefont {Krauss},
  \citenamefont {Roth}, \citenamefont {Alebrand}, \citenamefont {Steil},
  \citenamefont {Cinchetti}, \citenamefont {Aeschlimann},\ and\ \citenamefont
  {Schneider}}]{krauss09}%
  \BibitemOpen
  \bibfield  {author} {\bibinfo {author} {\bibfnamefont {M.}~\bibnamefont
  {Krauss}}, \bibinfo {author} {\bibfnamefont {T.}~\bibnamefont {Roth}},
  \bibinfo {author} {\bibfnamefont {S.}~\bibnamefont {Alebrand}}, \bibinfo
  {author} {\bibfnamefont {D.}~\bibnamefont {Steil}}, \bibinfo {author}
  {\bibfnamefont {M.}~\bibnamefont {Cinchetti}}, \bibinfo {author}
  {\bibfnamefont {M.}~\bibnamefont {Aeschlimann}}, \ and\ \bibinfo {author}
  {\bibfnamefont {H.~C.}\ \bibnamefont {Schneider}},\ }\href@noop {} {\bibfield
   {journal} {\bibinfo  {journal} {Phys. Rev. B}\ }\textbf {\bibinfo {volume}
  {80}},\ \bibinfo {pages} {180407(R)} (\bibinfo {year} {2009})}\BibitemShut
  {NoStop}%
\bibitem [{\citenamefont {Battiato}\ \emph {et~al.}(2010)\citenamefont
  {Battiato}, \citenamefont {Carva},\ and\ \citenamefont
  {Oppeneer}}]{battiato10}%
  \BibitemOpen
  \bibfield  {author} {\bibinfo {author} {\bibfnamefont {M.}~\bibnamefont
  {Battiato}}, \bibinfo {author} {\bibfnamefont {K.}~\bibnamefont {Carva}}, \
  and\ \bibinfo {author} {\bibfnamefont {P.~M.}\ \bibnamefont {Oppeneer}},\
  }\href@noop {} {\bibfield  {journal} {\bibinfo  {journal} {Phys. Rev. Lett.}\
  }\textbf {\bibinfo {volume} {105}},\ \bibinfo {pages} {027203} (\bibinfo
  {year} {2010})}\BibitemShut {NoStop}%
\bibitem [{\citenamefont {Battiato}\ \emph {et~al.}(2012)\citenamefont
  {Battiato}, \citenamefont {Carva},\ and\ \citenamefont
  {Oppeneer}}]{battiato12}%
  \BibitemOpen
  \bibfield  {author} {\bibinfo {author} {\bibfnamefont {M.}~\bibnamefont
  {Battiato}}, \bibinfo {author} {\bibfnamefont {K.}~\bibnamefont {Carva}}, \
  and\ \bibinfo {author} {\bibfnamefont {P.~M.}\ \bibnamefont {Oppeneer}},\
  }\href@noop {} {\bibfield  {journal} {\bibinfo  {journal} {Phys. Rev. B}\
  }\textbf {\bibinfo {volume} {86}},\ \bibinfo {pages} {024404} (\bibinfo
  {year} {2012})}\BibitemShut {NoStop}%
\bibitem [{\citenamefont {Zhang}\ \emph {et~al.}(2009)\citenamefont {Zhang},
  \citenamefont {H{\"u}bner}, \citenamefont {Lefkidis}, \citenamefont {Bai},\
  and\ \citenamefont {George}}]{zhang09}%
  \BibitemOpen
  \bibfield  {author} {\bibinfo {author} {\bibfnamefont {G.~P.}\ \bibnamefont
  {Zhang}}, \bibinfo {author} {\bibfnamefont {W.}~\bibnamefont {H{\"u}bner}},
  \bibinfo {author} {\bibfnamefont {G.}~\bibnamefont {Lefkidis}}, \bibinfo
  {author} {\bibfnamefont {Y.}~\bibnamefont {Bai}}, \ and\ \bibinfo {author}
  {\bibfnamefont {T.~F.}\ \bibnamefont {George}},\ }\href@noop {} {\bibfield
  {journal} {\bibinfo  {journal} {Nature Phys.}\ }\textbf {\bibinfo {volume}
  {5}},\ \bibinfo {pages} {499} (\bibinfo {year} {2009})}\BibitemShut {NoStop}%
\bibitem [{\citenamefont {Bigot}\ \emph {et~al.}(2009)\citenamefont {Bigot},
  \citenamefont {Vomir},\ and\ \citenamefont {Beaurepaire}}]{bigot09}%
  \BibitemOpen
  \bibfield  {author} {\bibinfo {author} {\bibfnamefont {J.-Y.}\ \bibnamefont
  {Bigot}}, \bibinfo {author} {\bibfnamefont {M.}~\bibnamefont {Vomir}}, \ and\
  \bibinfo {author} {\bibfnamefont {E.}~\bibnamefont {Beaurepaire}},\
  }\href@noop {} {\bibfield  {journal} {\bibinfo  {journal} {Nature Phys.}\
  }\textbf {\bibinfo {volume} {5}},\ \bibinfo {pages} {515} (\bibinfo {year}
  {2009})}\BibitemShut {NoStop}%
\bibitem [{\citenamefont {Atxitia}\ \emph {et~al.}(2010)\citenamefont
  {Atxitia}, \citenamefont {Chubykalo-Fesenko}, \citenamefont {Walowski},
  \citenamefont {Mann},\ and\ \citenamefont {M{\"u}nzenberg}}]{Atxitia10}%
  \BibitemOpen
  \bibfield  {author} {\bibinfo {author} {\bibfnamefont {U.}~\bibnamefont
  {Atxitia}}, \bibinfo {author} {\bibfnamefont {O.}~\bibnamefont
  {Chubykalo-Fesenko}}, \bibinfo {author} {\bibfnamefont {J.}~\bibnamefont
  {Walowski}}, \bibinfo {author} {\bibfnamefont {A.}~\bibnamefont {Mann}}, \
  and\ \bibinfo {author} {\bibfnamefont {M.}~\bibnamefont {M{\"u}nzenberg}},\
  }\href@noop {} {\bibfield  {journal} {\bibinfo  {journal} {Phys. Rev. B}\
  }\textbf {\bibinfo {volume} {81}},\ \bibinfo {pages} {174401} (\bibinfo
  {year} {2010})}\BibitemShut {NoStop}%
\bibitem [{\citenamefont {Ostler}\ \emph {et~al.}(2012)\citenamefont {Ostler},
  \citenamefont {Barker}, \citenamefont {Evans}, \citenamefont {Chantrell},
  \citenamefont {Atxitia}, \citenamefont {Chubykalo-Fesenko}, \citenamefont
  {El~Moussaoui}, \citenamefont {Le~Guyader}, \citenamefont {Mengotti},
  \citenamefont {Heyderman}, \citenamefont {Nolting}, \citenamefont
  {Tsukamoto}, \citenamefont {Itoh}, \citenamefont {Afanasiev}, \citenamefont
  {Ivanov}, \citenamefont {Kalashnikova}, \citenamefont {Vahaplar},
  \citenamefont {Mentink}, \citenamefont {Kirilyuk}, \citenamefont {Rasing},\
  and\ \citenamefont {Kimel}}]{Ostler12}%
  \BibitemOpen
  \bibfield  {author} {\bibinfo {author} {\bibfnamefont {T.~A.}\ \bibnamefont
  {Ostler}}, \bibinfo {author} {\bibfnamefont {J.}~\bibnamefont {Barker}},
  \bibinfo {author} {\bibfnamefont {R.~F.~L.}\ \bibnamefont {Evans}}, \bibinfo
  {author} {\bibfnamefont {R.~W.}\ \bibnamefont {Chantrell}}, \bibinfo {author}
  {\bibfnamefont {U.}~\bibnamefont {Atxitia}}, \bibinfo {author} {\bibfnamefont
  {O.}~\bibnamefont {Chubykalo-Fesenko}}, \bibinfo {author} {\bibfnamefont
  {S.}~\bibnamefont {El~Moussaoui}}, \bibinfo {author} {\bibfnamefont
  {L.}~\bibnamefont {Le~Guyader}}, \bibinfo {author} {\bibfnamefont
  {E.}~\bibnamefont {Mengotti}}, \bibinfo {author} {\bibfnamefont {L.~J.}\
  \bibnamefont {Heyderman}}, \bibinfo {author} {\bibfnamefont {F.}~\bibnamefont
  {Nolting}}, \bibinfo {author} {\bibfnamefont {A.}~\bibnamefont {Tsukamoto}},
  \bibinfo {author} {\bibfnamefont {A.}~\bibnamefont {Itoh}}, \bibinfo {author}
  {\bibfnamefont {D.}~\bibnamefont {Afanasiev}}, \bibinfo {author}
  {\bibfnamefont {B.~A.}\ \bibnamefont {Ivanov}}, \bibinfo {author}
  {\bibfnamefont {A.~M.}\ \bibnamefont {Kalashnikova}}, \bibinfo {author}
  {\bibfnamefont {K.}~\bibnamefont {Vahaplar}}, \bibinfo {author}
  {\bibfnamefont {J.}~\bibnamefont {Mentink}}, \bibinfo {author} {\bibfnamefont
  {A.}~\bibnamefont {Kirilyuk}}, \bibinfo {author} {\bibfnamefont {{\rm
  Th}.}~\bibnamefont {Rasing}}, \ and\ \bibinfo {author} {\bibfnamefont
  {A.~V.}\ \bibnamefont {Kimel}},\ }\href@noop {} {\bibfield  {journal}
  {\bibinfo  {journal} {Nature Commun.}\ }\textbf {\bibinfo {volume} {3}},\
  \bibinfo {pages} {666} (\bibinfo {year} {2012})}\BibitemShut {NoStop}%
\bibitem [{\citenamefont {Wienholdt}\ \emph {et~al.}(2013)\citenamefont
  {Wienholdt}, \citenamefont {Hinzke}, \citenamefont {Carva}, \citenamefont
  {Oppeneer},\ and\ \citenamefont {Nowak}}]{Wienholdt13}%
  \BibitemOpen
  \bibfield  {author} {\bibinfo {author} {\bibfnamefont {S.}~\bibnamefont
  {Wienholdt}}, \bibinfo {author} {\bibfnamefont {D.}~\bibnamefont {Hinzke}},
  \bibinfo {author} {\bibfnamefont {K.}~\bibnamefont {Carva}}, \bibinfo
  {author} {\bibfnamefont {P.~M.}\ \bibnamefont {Oppeneer}}, \ and\ \bibinfo
  {author} {\bibfnamefont {U.}~\bibnamefont {Nowak}},\ }\href {\doibase
  10.1103/PhysRevB.88.020406} {\bibfield  {journal} {\bibinfo  {journal} {Phys.
  Rev. B}\ }\textbf {\bibinfo {volume} {88}},\ \bibinfo {pages} {020406}
  (\bibinfo {year} {2013})}\BibitemShut {NoStop}%
\bibitem [{\citenamefont {Bar'yakhtar}\ \emph {et~al.}(2013)\citenamefont
  {Bar'yakhtar}, \citenamefont {Butrim},\ and\ \citenamefont
  {Ivanov}}]{Baryakhtar13}%
  \BibitemOpen
  \bibfield  {author} {\bibinfo {author} {\bibfnamefont {V.~G.}\ \bibnamefont
  {Bar'yakhtar}}, \bibinfo {author} {\bibfnamefont {V.~I.}\ \bibnamefont
  {Butrim}}, \ and\ \bibinfo {author} {\bibfnamefont {B.~A.}\ \bibnamefont
  {Ivanov}},\ }\href {\doibase 10.1134/S0021364013180057} {\bibfield  {journal}
  {\bibinfo  {journal} {JETP Lett.}\ }\textbf {\bibinfo {volume} {98}},\
  \bibinfo {pages} {289} (\bibinfo {year} {2013})}\BibitemShut {NoStop}%
\bibitem [{\citenamefont {Carva}\ \emph
  {et~al.}(2011{\natexlab{b}})\citenamefont {Carva}, \citenamefont {Battiato},\
  and\ \citenamefont {Oppeneer}}]{carva11}%
  \BibitemOpen
  \bibfield  {author} {\bibinfo {author} {\bibfnamefont {K.}~\bibnamefont
  {Carva}}, \bibinfo {author} {\bibfnamefont {M.}~\bibnamefont {Battiato}}, \
  and\ \bibinfo {author} {\bibfnamefont {P.~M.}\ \bibnamefont {Oppeneer}},\
  }\href@noop {} {\bibfield  {journal} {\bibinfo  {journal} {Phys. Rev. Lett.}\
  }\textbf {\bibinfo {volume} {107}},\ \bibinfo {pages} {207201} (\bibinfo
  {year} {2011}{\natexlab{b}})}\BibitemShut {NoStop}%
\bibitem [{\citenamefont {Essert}\ and\ \citenamefont
  {Schneider}(2011)}]{essert11}%
  \BibitemOpen
  \bibfield  {author} {\bibinfo {author} {\bibfnamefont {S.}~\bibnamefont
  {Essert}}\ and\ \bibinfo {author} {\bibfnamefont {H.~C.}\ \bibnamefont
  {Schneider}},\ }\href@noop {} {\bibfield  {journal} {\bibinfo  {journal}
  {Phys. Rev. B}\ }\textbf {\bibinfo {volume} {84}},\ \bibinfo {pages} {224405}
  (\bibinfo {year} {2011})}\BibitemShut {NoStop}%
\bibitem [{\citenamefont {Carva}\ \emph {et~al.}(2013)\citenamefont {Carva},
  \citenamefont {Battiato}, \citenamefont {Legut},\ and\ \citenamefont
  {Oppeneer}}]{carva13}%
  \BibitemOpen
  \bibfield  {author} {\bibinfo {author} {\bibfnamefont {K.}~\bibnamefont
  {Carva}}, \bibinfo {author} {\bibfnamefont {M.}~\bibnamefont {Battiato}},
  \bibinfo {author} {\bibfnamefont {D.}~\bibnamefont {Legut}}, \ and\ \bibinfo
  {author} {\bibfnamefont {P.~M.}\ \bibnamefont {Oppeneer}},\ }\href@noop {}
  {\bibfield  {journal} {\bibinfo  {journal} {Phys. Rev. B}\ }\textbf {\bibinfo
  {volume} {87}},\ \bibinfo {pages} {184425} (\bibinfo {year}
  {2013})}\BibitemShut {NoStop}%
\bibitem [{\citenamefont {Illg}\ \emph {et~al.}(2013)\citenamefont {Illg},
  \citenamefont {Haag},\ and\ \citenamefont {F\"ahnle}}]{illg13}%
  \BibitemOpen
  \bibfield  {author} {\bibinfo {author} {\bibfnamefont {C.}~\bibnamefont
  {Illg}}, \bibinfo {author} {\bibfnamefont {M.}~\bibnamefont {Haag}}, \ and\
  \bibinfo {author} {\bibfnamefont {M.}~\bibnamefont {F\"ahnle}},\ }\href
  {\doibase 10.1103/PhysRevB.88.214404} {\bibfield  {journal} {\bibinfo
  {journal} {Phys. Rev. B}\ }\textbf {\bibinfo {volume} {88}},\ \bibinfo
  {pages} {214404} (\bibinfo {year} {2013})}\BibitemShut {NoStop}%
\bibitem [{\citenamefont {Schellekens}\ and\ \citenamefont
  {Koopmans}(2013)}]{schellekens13}%
  \BibitemOpen
  \bibfield  {author} {\bibinfo {author} {\bibfnamefont {A.~J.}\ \bibnamefont
  {Schellekens}}\ and\ \bibinfo {author} {\bibfnamefont {B.}~\bibnamefont
  {Koopmans}},\ }\href@noop {} {\bibfield  {journal} {\bibinfo  {journal}
  {Phys. Rev. Lett.}\ }\textbf {\bibinfo {volume} {110}},\ \bibinfo {pages}
  {217204} (\bibinfo {year} {2013})}\BibitemShut {NoStop}%
\bibitem [{\citenamefont {Mueller}\ \emph {et~al.}(2013)\citenamefont
  {Mueller}, \citenamefont {Baral}, \citenamefont {Vollmar}, \citenamefont
  {Cinchetti}, \citenamefont {Aeschlimann}, \citenamefont {Schneider},\ and\
  \citenamefont {Rethfeld}}]{mueller13}%
  \BibitemOpen
  \bibfield  {author} {\bibinfo {author} {\bibfnamefont {B.~Y.}\ \bibnamefont
  {Mueller}}, \bibinfo {author} {\bibfnamefont {A.}~\bibnamefont {Baral}},
  \bibinfo {author} {\bibfnamefont {S.}~\bibnamefont {Vollmar}}, \bibinfo
  {author} {\bibfnamefont {M.}~\bibnamefont {Cinchetti}}, \bibinfo {author}
  {\bibfnamefont {M.}~\bibnamefont {Aeschlimann}}, \bibinfo {author}
  {\bibfnamefont {H.~C.}\ \bibnamefont {Schneider}}, \ and\ \bibinfo {author}
  {\bibfnamefont {B.}~\bibnamefont {Rethfeld}},\ }\href@noop {} {\bibfield
  {journal} {\bibinfo  {journal} {Phys. Rev. Lett.}\ }\textbf {\bibinfo
  {volume} {111}},\ \bibinfo {pages} {167204} (\bibinfo {year}
  {2013})}\BibitemShut {NoStop}%
\bibitem [{\citenamefont {Haag}\ \emph {et~al.}(2014)\citenamefont {Haag},
  \citenamefont {Illg},\ and\ \citenamefont {F\"ahnle}}]{haag14}%
  \BibitemOpen
  \bibfield  {author} {\bibinfo {author} {\bibfnamefont {M.}~\bibnamefont
  {Haag}}, \bibinfo {author} {\bibfnamefont {C.}~\bibnamefont {Illg}}, \ and\
  \bibinfo {author} {\bibfnamefont {M.}~\bibnamefont {F\"ahnle}},\ }\href
  {\doibase 10.1103/PhysRevB.90.014417} {\bibfield  {journal} {\bibinfo
  {journal} {Phys. Rev. B}\ }\textbf {\bibinfo {volume} {90}},\ \bibinfo
  {pages} {014417} (\bibinfo {year} {2014})}\BibitemShut {NoStop}%
\bibitem [{\citenamefont {Dixit}\ \emph {et~al.}(2013)\citenamefont {Dixit},
  \citenamefont {Hinschberger}, \citenamefont {Zamanian}, \citenamefont
  {Manfredi},\ and\ \citenamefont {Hervieux}}]{dixit13}%
  \BibitemOpen
  \bibfield  {author} {\bibinfo {author} {\bibfnamefont {A.}~\bibnamefont
  {Dixit}}, \bibinfo {author} {\bibfnamefont {Y.}~\bibnamefont {Hinschberger}},
  \bibinfo {author} {\bibfnamefont {J.}~\bibnamefont {Zamanian}}, \bibinfo
  {author} {\bibfnamefont {G.}~\bibnamefont {Manfredi}}, \ and\ \bibinfo
  {author} {\bibfnamefont {P.-A.}\ \bibnamefont {Hervieux}},\ }\href {\doibase
  10.1103/PhysRevA.88.032117} {\bibfield  {journal} {\bibinfo  {journal} {Phys.
  Rev. A}\ }\textbf {\bibinfo {volume} {88}},\ \bibinfo {pages} {032117}
  (\bibinfo {year} {2013})}\BibitemShut {NoStop}%
\bibitem [{\citenamefont {Hinschberger}\ and\ \citenamefont
  {Hervieux}(2013)}]{hinschberger13}%
  \BibitemOpen
  \bibfield  {author} {\bibinfo {author} {\bibfnamefont {Y.}~\bibnamefont
  {Hinschberger}}\ and\ \bibinfo {author} {\bibfnamefont {P.-A.}\ \bibnamefont
  {Hervieux}},\ }\href {\doibase 10.1103/PhysRevB.88.134413} {\bibfield
  {journal} {\bibinfo  {journal} {Phys. Rev. B}\ }\textbf {\bibinfo {volume}
  {88}},\ \bibinfo {pages} {134413} (\bibinfo {year} {2013})}\BibitemShut
  {NoStop}%
\bibitem [{\citenamefont {Vonesch}\ and\ \citenamefont
  {Bigot}(2012)}]{vonesch12}%
  \BibitemOpen
  \bibfield  {author} {\bibinfo {author} {\bibfnamefont {H.}~\bibnamefont
  {Vonesch}}\ and\ \bibinfo {author} {\bibfnamefont {J.-Y.}\ \bibnamefont
  {Bigot}},\ }\href {\doibase 10.1103/PhysRevB.85.180407} {\bibfield  {journal}
  {\bibinfo  {journal} {Phys. Rev. B}\ }\textbf {\bibinfo {volume} {85}},\
  \bibinfo {pages} {180407} (\bibinfo {year} {2012})}\BibitemShut {NoStop}%
\bibitem [{\citenamefont {MacDonald}\ and\ \citenamefont
  {Vosko}(1979)}]{macdonald79}%
  \BibitemOpen
  \bibfield  {author} {\bibinfo {author} {\bibfnamefont {A.~H.}\ \bibnamefont
  {MacDonald}}\ and\ \bibinfo {author} {\bibfnamefont {S.~H.}\ \bibnamefont
  {Vosko}},\ }\href@noop {} {\bibfield  {journal} {\bibinfo  {journal} {J.
  Phys. C: Solid State Phys.}\ }\textbf {\bibinfo {volume} {12}},\ \bibinfo
  {pages} {2977} (\bibinfo {year} {1979})}\BibitemShut {NoStop}%
\bibitem [{\citenamefont {Eschrig}\ and\ \citenamefont
  {Servedio}(1999)}]{eschrig99}%
  \BibitemOpen
  \bibfield  {author} {\bibinfo {author} {\bibfnamefont {H.}~\bibnamefont
  {Eschrig}}\ and\ \bibinfo {author} {\bibfnamefont {V.~D.~P.}\ \bibnamefont
  {Servedio}},\ }\href@noop {} {\bibfield  {journal} {\bibinfo  {journal} {J.
  Comput. Chem.}\ }\textbf {\bibinfo {volume} {20}},\ \bibinfo {pages} {23}
  (\bibinfo {year} {1999})}\BibitemShut {NoStop}%
\bibitem [{\citenamefont {Greiner}(2000)}]{greiner00}%
  \BibitemOpen
  \bibfield  {author} {\bibinfo {author} {\bibfnamefont {W.}~\bibnamefont
  {Greiner}},\ }\href@noop {} {\emph {\bibinfo {title} {Relativistic quantum
  mechanics. Wave equations}}}\ (\bibinfo  {publisher} {Springer, Berlin},\
  \bibinfo {year} {2000})\BibitemShut {NoStop}%
\bibitem [{\citenamefont {Kraft}\ \emph {et~al.}(1995)\citenamefont {Kraft},
  \citenamefont {Oppeneer}, \citenamefont {Antonov},\ and\ \citenamefont
  {Eschrig}}]{kraft95}%
  \BibitemOpen
  \bibfield  {author} {\bibinfo {author} {\bibfnamefont {T.}~\bibnamefont
  {Kraft}}, \bibinfo {author} {\bibfnamefont {P.~M.}\ \bibnamefont {Oppeneer}},
  \bibinfo {author} {\bibfnamefont {V.~N.}\ \bibnamefont {Antonov}}, \ and\
  \bibinfo {author} {\bibfnamefont {H.}~\bibnamefont {Eschrig}},\ }\href@noop
  {} {\bibfield  {journal} {\bibinfo  {journal} {Phys. Rev. B}\ }\textbf
  {\bibinfo {volume} {52}},\ \bibinfo {pages} {3561} (\bibinfo {year}
  {1995})}\BibitemShut {NoStop}%
\bibitem [{\citenamefont {Foldy}\ and\ \citenamefont
  {Wouthuysen}(1950)}]{foldy50}%
  \BibitemOpen
  \bibfield  {author} {\bibinfo {author} {\bibfnamefont {L.~L.}\ \bibnamefont
  {Foldy}}\ and\ \bibinfo {author} {\bibfnamefont {S.~A.}\ \bibnamefont
  {Wouthuysen}},\ }\href {\doibase 10.1103/PhysRev.78.29} {\bibfield  {journal}
  {\bibinfo  {journal} {Phys. Rev.}\ }\textbf {\bibinfo {volume} {78}},\
  \bibinfo {pages} {29} (\bibinfo {year} {1950})}\BibitemShut {NoStop}%
\bibitem [{\citenamefont {Cr\'epieux}\ and\ \citenamefont
  {Bruno}(2001)}]{crepieux01}%
  \BibitemOpen
  \bibfield  {author} {\bibinfo {author} {\bibfnamefont {A.}~\bibnamefont
  {Cr\'epieux}}\ and\ \bibinfo {author} {\bibfnamefont {P.}~\bibnamefont
  {Bruno}},\ }\href {\doibase 10.1103/PhysRevB.64.094434} {\bibfield  {journal}
  {\bibinfo  {journal} {Phys. Rev. B}\ }\textbf {\bibinfo {volume} {64}},\
  \bibinfo {pages} {094434} (\bibinfo {year} {2001})}\BibitemShut {NoStop}%
\bibitem [{\citenamefont {Oppeneer}(2001)}]{oppeneer01}%
  \BibitemOpen
  \bibfield  {author} {\bibinfo {author} {\bibfnamefont {P.~M.}\ \bibnamefont
  {Oppeneer}},\ }in\ \href@noop {} {\emph {\bibinfo {booktitle} {Handbook of
  Magnetic Materials}}},\ Vol.~\bibinfo {volume} {13},\ \bibinfo {editor}
  {edited by\ \bibinfo {editor} {\bibfnamefont {K.~H.~J.}\ \bibnamefont
  {Buschow}}}\ (\bibinfo  {publisher} {Elsevier, Amsterdam},\ \bibinfo {year}
  {2001})\ pp.\ \bibinfo {pages} {229 -- 422}\BibitemShut {NoStop}%
\bibitem [{\citenamefont {Landau}\ and\ \citenamefont
  {Lifshitz}(1981)}]{landau81}%
  \BibitemOpen
  \bibfield  {author} {\bibinfo {author} {\bibfnamefont {L.~D.}\ \bibnamefont
  {Landau}}\ and\ \bibinfo {author} {\bibfnamefont {L.~M.}\ \bibnamefont
  {Lifshitz}},\ }\href {http://www.worldcat.org/isbn/0750635398} {\emph
  {\bibinfo {title} {Quantum Mechanics: Non-Relativistic Theory}}},\ \bibinfo
  {edition} {3rd}\ ed.,\ Vol.~\bibinfo {volume} {3}\ (\bibinfo  {publisher}
  {Butterworth-Heinemann, Oxford},\ \bibinfo {year} {1981})\BibitemShut
  {NoStop}%
\bibitem [{\citenamefont {Nowakowski}(1999)}]{nowakowski99}%
  \BibitemOpen
  \bibfield  {author} {\bibinfo {author} {\bibfnamefont {M.}~\bibnamefont
  {Nowakowski}},\ }\href@noop {} {\bibfield  {journal} {\bibinfo  {journal}
  {Am. J. Phys.}\ }\textbf {\bibinfo {volume} {67}},\ \bibinfo {pages} {916}
  (\bibinfo {year} {1999})}\BibitemShut {NoStop}%
\bibitem [{\citenamefont {Husz{\'a}r}(1967)}]{huszar67}%
  \BibitemOpen
  \bibfield  {author} {\bibinfo {author} {\bibfnamefont {M.}~\bibnamefont
  {Husz{\'a}r}},\ }\href@noop {} {\bibfield  {journal} {\bibinfo  {journal}
  {Acta Phys. Acad. Scien. Hungaricae}\ }\textbf {\bibinfo {volume} {23}},\
  \bibinfo {pages} {225} (\bibinfo {year} {1967})}\BibitemShut {NoStop}%
\bibitem [{\citenamefont {Ullrich}(2011)}]{ullrich11}%
  \BibitemOpen
  \bibfield  {author} {\bibinfo {author} {\bibfnamefont {C.~A.}\ \bibnamefont
  {Ullrich}},\ }\href@noop {} {\emph {\bibinfo {title} {Time-dependent
  density-functional theory: Concepts and applications}}}\ (\bibinfo
  {publisher} {Oxford University Press, Oxford},\ \bibinfo {year}
  {2011})\BibitemShut {NoStop}%
\bibitem [{\citenamefont {Marques}\ \emph {et~al.}(2012)\citenamefont
  {Marques}, \citenamefont {Maitra}, \citenamefont {Nogueira}, \citenamefont
  {Gross},\ and\ \citenamefont {Rubio}}]{marques12}%
  \BibitemOpen
  \bibinfo {editor} {\bibfnamefont {M.~A.}\ \bibnamefont {Marques}}, \bibinfo
  {editor} {\bibfnamefont {N.~T.}\ \bibnamefont {Maitra}}, \bibinfo {editor}
  {\bibfnamefont {F.~M.}\ \bibnamefont {Nogueira}}, \bibinfo {editor}
  {\bibfnamefont {E.~K.~U.}\ \bibnamefont {Gross}}, \ and\ \bibinfo {editor}
  {\bibfnamefont {A.}~\bibnamefont {Rubio}},\ eds.,\ \href@noop {} {\emph
  {\bibinfo {title} {Fundamentals of Time-Dependent Density Functional
  Theory}}},\ \bibinfo {series} {Lecture Notes in Physics}, Vol.\ \bibinfo
  {volume} {837}\ (\bibinfo  {publisher} {Springer, Heidelberg},\ \bibinfo
  {year} {2012})\BibitemShut {NoStop}%
\bibitem [{\citenamefont {Koopmans}\ \emph {et~al.}(2000)\citenamefont
  {Koopmans}, \citenamefont {van Kampen}, \citenamefont {Kohlhepp},\ and\
  \citenamefont {de~Jonge}}]{koopmans00}%
  \BibitemOpen
  \bibfield  {author} {\bibinfo {author} {\bibfnamefont {B.}~\bibnamefont
  {Koopmans}}, \bibinfo {author} {\bibfnamefont {M.}~\bibnamefont {van
  Kampen}}, \bibinfo {author} {\bibfnamefont {J.~T.}\ \bibnamefont {Kohlhepp}},
  \ and\ \bibinfo {author} {\bibfnamefont {W.~J.~M.}\ \bibnamefont
  {de~Jonge}},\ }\href {\doibase 10.1103/PhysRevLett.85.844} {\bibfield
  {journal} {\bibinfo  {journal} {Phys. Rev. Lett.}\ }\textbf {\bibinfo
  {volume} {85}},\ \bibinfo {pages} {844} (\bibinfo {year} {2000})}\BibitemShut
  {NoStop}%
\bibitem [{\citenamefont {Oppeneer}\ \emph
  {et~al.}(1992{\natexlab{a}})\citenamefont {Oppeneer}, \citenamefont {Maurer},
  \citenamefont {Sticht},\ and\ \citenamefont {K{\"u}bler}}]{oppeneer1992ab}%
  \BibitemOpen
  \bibfield  {author} {\bibinfo {author} {\bibfnamefont {P.~M.}\ \bibnamefont
  {Oppeneer}}, \bibinfo {author} {\bibfnamefont {T.}~\bibnamefont {Maurer}},
  \bibinfo {author} {\bibfnamefont {J.}~\bibnamefont {Sticht}}, \ and\ \bibinfo
  {author} {\bibfnamefont {J.}~\bibnamefont {K{\"u}bler}},\ }\href@noop {}
  {\bibfield  {journal} {\bibinfo  {journal} {Phys. Rev. B}\ }\textbf {\bibinfo
  {volume} {45}},\ \bibinfo {pages} {10924} (\bibinfo {year}
  {1992}{\natexlab{a}})}\BibitemShut {NoStop}%
\bibitem [{\citenamefont {Argyres}(1955)}]{Argyres55}%
  \BibitemOpen
  \bibfield  {author} {\bibinfo {author} {\bibfnamefont {P.~N.}\ \bibnamefont
  {Argyres}},\ }\href@noop {} {\bibfield  {journal} {\bibinfo  {journal} {Phys.
  Rev.}\ }\textbf {\bibinfo {volume} {97}},\ \bibinfo {pages} {334} (\bibinfo
  {year} {1955})}\BibitemShut {NoStop}%
\bibitem [{\citenamefont {Oppeneer}\ \emph
  {et~al.}(1992{\natexlab{b}})\citenamefont {Oppeneer}, \citenamefont {Sticht},
  \citenamefont {Maurer},\ and\ \citenamefont {K{\"u}bler}}]{oppeneer1992abso}%
  \BibitemOpen
  \bibfield  {author} {\bibinfo {author} {\bibfnamefont {P.~M.}\ \bibnamefont
  {Oppeneer}}, \bibinfo {author} {\bibfnamefont {J.}~\bibnamefont {Sticht}},
  \bibinfo {author} {\bibfnamefont {T.}~\bibnamefont {Maurer}}, \ and\ \bibinfo
  {author} {\bibfnamefont {J.}~\bibnamefont {K{\"u}bler}},\ }\href@noop {}
  {\bibfield  {journal} {\bibinfo  {journal} {Z. Physik B}\ }\textbf {\bibinfo
  {volume} {88}},\ \bibinfo {pages} {309} (\bibinfo {year}
  {1992}{\natexlab{b}})}\BibitemShut {NoStop}%
\bibitem [{\citenamefont {Williams}\ \emph {et~al.}(1979)\citenamefont
  {Williams}, \citenamefont {K\"ubler},\ and\ \citenamefont
  {Gelatt}}]{williams79}%
  \BibitemOpen
  \bibfield  {author} {\bibinfo {author} {\bibfnamefont {A.~R.}\ \bibnamefont
  {Williams}}, \bibinfo {author} {\bibfnamefont {J.}~\bibnamefont {K\"ubler}},
  \ and\ \bibinfo {author} {\bibfnamefont {C.~D.}\ \bibnamefont {Gelatt}},\
  }\href {\doibase 10.1103/PhysRevB.19.6094} {\bibfield  {journal} {\bibinfo
  {journal} {Phys. Rev. B}\ }\textbf {\bibinfo {volume} {19}},\ \bibinfo
  {pages} {6094} (\bibinfo {year} {1979})}\BibitemShut {NoStop}%
\bibitem [{\citenamefont {Guidoni}\ \emph {et~al.}(2002)\citenamefont
  {Guidoni}, \citenamefont {Beaurepaire},\ and\ \citenamefont
  {Bigot}}]{guidoni02}%
  \BibitemOpen
  \bibfield  {author} {\bibinfo {author} {\bibfnamefont {L.}~\bibnamefont
  {Guidoni}}, \bibinfo {author} {\bibfnamefont {E.}~\bibnamefont
  {Beaurepaire}}, \ and\ \bibinfo {author} {\bibfnamefont {J.-Y.}\ \bibnamefont
  {Bigot}},\ }\href@noop {} {\bibfield  {journal} {\bibinfo  {journal} {Phys.
  Rev. Lett.}\ }\textbf {\bibinfo {volume} {{89}}} (\bibinfo {year}
  {{2002}})}\BibitemShut {NoStop}%
\bibitem [{\citenamefont {Oppeneer}\ and\ \citenamefont
  {Liebsch}(2004)}]{oppeneer04}%
  \BibitemOpen
  \bibfield  {author} {\bibinfo {author} {\bibfnamefont {P.~M.}\ \bibnamefont
  {Oppeneer}}\ and\ \bibinfo {author} {\bibfnamefont {A.}~\bibnamefont
  {Liebsch}},\ }\href@noop {} {\bibfield  {journal} {\bibinfo  {journal} {J.
  Phys.: Condens. Matter}\ }\textbf {\bibinfo {volume} {16}},\ \bibinfo {pages}
  {5519} (\bibinfo {year} {2004})}\BibitemShut {NoStop}%
\bibitem [{\citenamefont {Krieger}\ \emph {et~al.}(2014)\citenamefont
  {Krieger}, \citenamefont {Dewhurst}, \citenamefont {Elliott}, \citenamefont
  {Sharma},\ and\ \citenamefont {Gross}}]{krieger14}%
  \BibitemOpen
  \bibfield  {author} {\bibinfo {author} {\bibfnamefont {K.}~\bibnamefont
  {Krieger}}, \bibinfo {author} {\bibfnamefont {J.~K.}\ \bibnamefont
  {Dewhurst}}, \bibinfo {author} {\bibfnamefont {P.}~\bibnamefont {Elliott}},
  \bibinfo {author} {\bibfnamefont {S.}~\bibnamefont {Sharma}}, \ and\ \bibinfo
  {author} {\bibfnamefont {E.~K.~U.}\ \bibnamefont {Gross}},\ }\href@noop {}
  {\bibfield  {journal} {\bibinfo  {journal} {cond-mat arXiv:1406.6607v1}\ }
  (\bibinfo {year} {2014})}\BibitemShut {NoStop}%
\bibitem [{\citenamefont {Blundell}(2001)}]{blundell01}%
  \BibitemOpen
  \bibfield  {author} {\bibinfo {author} {\bibfnamefont {S.}~\bibnamefont
  {Blundell}},\ }\href
  {http://www.amazon.com/exec/obidos/redirect?tag=citeulike07-20\&path=ASIN/0198505914}
  {\emph {\bibinfo {title} {{Magnetism in Condensed Matter}}}}\ (\bibinfo
  {publisher} {Oxford University Press, Oxford},\ \bibinfo {year}
  {2001})\BibitemShut {NoStop}%
\bibitem [{\citenamefont {Eschrig}(1996)}]{eschrig96}%
  \BibitemOpen
  \bibfield  {author} {\bibinfo {author} {\bibfnamefont {H.}~\bibnamefont
  {Eschrig}},\ }\href@noop {} {\emph {\bibinfo {title} {The Fundamentals of
  Density Functional Theory}}}\ (\bibinfo  {publisher} {Teubner Verlag,
  Leipzig},\ \bibinfo {year} {1996})\BibitemShut {NoStop}%
\bibitem [{\citenamefont {Callaway}(1974)}]{callaway74}%
  \BibitemOpen
  \bibfield  {author} {\bibinfo {author} {\bibfnamefont {J.}~\bibnamefont
  {Callaway}},\ }\href {\doibase
  http://dx.doi.org/10.1016/B978-0-12-155202-2.50008-5} {\emph {\bibinfo
  {title} {Quantum Theory of the Solid State}}},\ \bibinfo {edition} {2nd}\
  ed.\ (\bibinfo  {publisher} {Academic Press, San Diego},\ \bibinfo {year}
  {1974})\ pp.\ \bibinfo {pages} {465 -- 572}\BibitemShut {NoStop}%
\end{thebibliography}
%merlin.mbs apsrev4-1.bst 2010-07-25 4.21a (PWD, AO, DPC) hacked
%Control: key (0)
%Control: author (72) initials jnrlst
%Control: editor formatted (1) identically to author
%Control: production of article title (-1) disabled
%Control: page (0) single
%Control: year (1) truncated
%Control: production of eprint (0) enabled
%

\end{document}